\documentclass[iop, revtex4]{emulateapj}

\usepackage{graphicx, natbib, bm, color}

\usepackage[backref,breaklinks,colorlinks,citecolor=blue]{hyperref}
\usepackage[all]{hypcap}

\newcommand{\noopsort}[1]{}
\newcommand{\yg}{\texttt{Yggdrasil}}
\newcommand{\cs}{\texttt{cluster\_slug}}
\newcommand{\slug}{\texttt{slug}}
\newcommand{\vecx}{\boldsymbol{x}}

\definecolor{darkred}{rgb}{0.7,0,0}
\newcommand{\red}[1]{{#1}}

\begin{document}

\title{Star Cluster Properties in Two LEGUS Galaxies Computed with Stochastic Stellar Population Synthesis Models}

\shorttitle{Stochastic Star Cluster Models in LEGUS}
\shortauthors{Krumholz et al.}

\author{Mark R.~Krumholz\altaffilmark{1}, Angela Adamo\altaffilmark{2}, Michele Fumagalli\altaffilmark{3}, Aida Wofford\altaffilmark{4}, Daniela Calzetti\altaffilmark{5}, Janice C.~Lee\altaffilmark{6, 7}, Bradley C.~Whitmore\altaffilmark{6}, Stacey N.~Bright\altaffilmark{6}, Kathryn Grasha\altaffilmark{5}, Dimitrios A.~Gouliermis\altaffilmark{8, 9}, Hwihyun Kim\altaffilmark{10,11}, Preethi Nair\altaffilmark{12}, Jenna E.~Ryon\altaffilmark{13}, Linda J.~Smith\altaffilmark{14}, David Thilker\altaffilmark{15}, Leonardo Ubeda\altaffilmark{6}, and Erik Zackrisson\altaffilmark{16}}

\altaffiltext{1}{Department of Astronomy and Astrophysics,
         University of California, Santa Cruz, CA 95064, USA;
         mkrumhol@ucsc.edu}
\altaffiltext{2}{Department of Astronomy, Oskar Klein Centre, Stockholm University, SE-10691 Stockholm, Sweden; adamo@astro.su.se}
\altaffiltext{3}{Institute for Computational Cosmology and Centre for Extragalactic Astronomy, Department of Physics, Durham University, South Road, Durham, DH1 3LE, UK}
\altaffiltext{4}{Institut d'Astrophysique de Paris, 98bis Boulevard Arago, 75014 Paris, France}
\altaffiltext{5}{Department of Astronomy, University of Massachusetts -- Amherst, Amherst, MA, USA}
\altaffiltext{6}{Space Telescope Science Institute, Baltimore, MD, USA}
\altaffiltext{7}{Visiting Astronomer, Spitzer Science Center, California Institute of Technology, Pasadena, CA, USA}
\altaffiltext{8}{Centre for Astronomy, Institute for Theoretical Astrophysics, University of Heidelberg, Heidelberg, Germany}
\altaffiltext{9}{Max Planck Institute for Astronomy, Heidelberg, Germany}
\altaffiltext{10}{Korea Astronomy and Space Science Institute, Daejeon, Republic of Korea}
\altaffiltext{11}{Department of Astronomy, The University of Texas at Austin, Austin, TX, USA}
\altaffiltext{12}{Department of Physics and Astronomy, University of Alabama, Tuscaloosa, AL, USA}
\altaffiltext{13}{Department of Astronomy, University of Wisconsin - Madison, Madison, WI, USA}
\altaffiltext{14}{European Space Agency/Space Telescope Science Institute, Baltimore, MD, USA}
\altaffiltext{15}{Department of Physics and Astronomy, The Johns Hopkins University, Baltimore, MD, USA}
\altaffiltext{16}{Department of Physics and Astronomy, Uppsala University, Uppsala, Sweden}

\begin{abstract}
We investigate a novel Bayesian analysis method, based on the Stochastically Lighting Up Galaxies (\slug) code, to derive the masses, ages, and extinctions of star clusters from integrated light photometry. Unlike many analysis methods, \slug~correctly accounts for incomplete IMF sampling, and returns full posterior probability distributions rather than simply probability maxima. We apply our technique to 621 visually-confirmed clusters in two nearby galaxies, NGC\,628 and NGC\,7793, that are part of the Legacy Extragalactic UV Survey (LEGUS). LEGUS provides \textit{Hubble Space Telescope} photometry in the NUV, U, B, V, and I bands. We analyze the sensitivity of the derived cluster properties to choices of prior probability distribution, evolutionary tracks, IMF, metallicity, treatment of nebular emission, and extinction curve. We find that \slug's results for individual clusters are insensitive to most of these choices, but that the posterior probability distributions we derive are often quite broad, and sometimes multi-peaked and quite sensitive to the choice of priors. In contrast, the properties of the cluster population as a whole are relatively robust against all of these choices. We also compare our results from \slug~to those derived with a conventional non-stochastic fitting code, \yg. We show that \slug's stochastic models are generally a better fit to the observations than the deterministic ones used by \yg. However, the overall properties of the cluster populations recovered by both codes are qualitatively similar.
\end{abstract}


\slugcomment{Accepted for publication in The Astrophysical Journal \today}

\section{Introduction}

Star clusters represent a scale of star formation intermediate between individual stars and entire galaxies. They likely represent the gravitationally-bound peaks of a continuous distribution of star formation across length and time scales. For this reason, the study of their properties, and any variation in those properties as a function of galactic environment, is critical to an overall theoretical understanding of the star formation process (see \citealt{portegies-zwart10a}, \citealt{longmore14a}, and \citealt{krumholz14c} for recent reviews).

While the study of star clusters is an old topic, except in the Milky Way and the very closest external galaxies, much of this work has by necessity focused on the relatively massive clusters, $\sim 10^{3.5}$ $M_\odot$ or more \citep[e.g.][]{zhang99b, larsen00a, bik03a, de-grijs03a, fall05a, goodwin06a, larsen09a}. This has created significant challenges for theoretical interpretation, for two reasons. The first is simply statistics: as discussed in more detail below, observed star cluster mass functions appear to be well-fit by powerlaws of the form $dN/dM \propto M^{-2}$ \citep[e.g.,][]{williams97a, zhang99a, larsen02a, bik03a, de-grijs03a, goddard10a, bastian11a, fall12a, fouesneau12a}. Such a mass function implies that clusters in the mass range $10^{2.5} - 10^{3.5}$ $M_\odot$ are nearly ten times as numerous as those in the range $\sim 10^{3.5} - 10^{5.5}$ $M_\odot$ that is typically studied, so a sample consisting only of massive clusters necessarily has much less statistical power than one probing to lower masses. The second is that theoretical models for the evolution and disruption of star clusters, either due to stellar feedback or due to environmental influence, make some of their strongest predictions for the effects of cluster mass at masses below this range \citep[e.g.][]{lamers05a, parmentier08a, fall10a, kruijssen12a, kruijssen12b}.

In the last five years a number of groups have made a concerted effort to extend the star cluster data set to lower masses and a broader range of galactic environments, a necessary step if we are to differentiate between models. Doing so is challenging both observationally and theoretically. Observationally, ground-based studies probing to lower masses are mostly limited by resolution to the Milky Way \citep{borissova11a} and the Magellanic Clouds \citep{hunter03a, rafelski05a}. Extending this sample is one of the primary goals of two ongoing \textit{Hubble Space Telescope} programs: the Panchromatic Hubble Andromeda Treasury (PHAT; \citealt{dalcanton12a}) and the Legacy Extra-Galactic UV Survey (LEGUS; \citealt{calzetti15a}). The former focuses on M31 to extreme depth, providing an extensive and very complete cluster catalog, but for only a single galaxy. The latter targets 50 nearby galaxies chosen from across the star-formiong portion of the Hubble sequence, providing data on a much greater range of galactic environments, but with less depth. A preliminary cluster catalog for PHAT is now available \citep{johnson12a}, and catalogs for the LEGUS galaxies will be released in the future (Adamo et al., 2015, in preparation).

The theoretical challenge when working with these extended cluster samples is that deriving masses and other properties for small clusters is non-trivial because they do not fully sample the stellar initial mass function \citep[e.g.,][]{cervino03a, cervino04a, cervino06a, maiz-apellaniz09a}. Traditional methods of determining star cluster properties by isochrone fitting therefore begin to fail, because at such small masses the relationship between clusters' photometric properties and their mass, age, and other physical properties is non-deterministic. One approach to dealing with this problem is to resolve at least the most massive members of the cluster and determine their properties star-by-star using a color-magnitude diagram \citep[CMD,][]{beerman12a}. However, for young open clusters where the stellar surface density is high and confusion is significant, this requires extreme resolution and depth in the observations, and thus is not yet practical as a method for obtaining large cluster samples across a wide range of galactic environments.\footnote{To our knowledge the largest-scale application of this technique to appear in the literature to date is in \citet{fouesneau14a}, who use PHAT photometry to obtain CMD-based ages for 100 clusters in M51. However, this paper presents the CMD-based results only for the purposes of comparison with those based on integrated photometry, and does not contain the CMD analysis itself, which is deferred to an as-yet unpublished paper.} The alternative approach is to develop new statistical techniques that can cope with partial sampling, and this has motivated the development of three major stochastic stellar population synthesis and analysis codes: \texttt{MASSclean} \citep{popescu09a, popescu10a, popescu10b}, a stochastic version of \texttt{pegase} \citep{fouesneau10a, fouesneau12a, fouesneau14a}, and \slug~\citep{da-silva12a, da-silva14b, krumholz15b}. While these codes use slightly different statistical techniques, they all attempt to solve essentially the same problem: given a set of observed photometric properties for a star cluster, what should we infer about the probability distribution for its mass, age, extinction, or other physical properties? 

In this paper we use \slug, the Stochastically Lighting Up Galaxies code, and its post-processing tool for analysis of star cluster properties, \cs, to analyze an initial sample of clusters from the Legacy Extragalactic Galaxy Ultraviolet Survey (LEGUS; \citealt{calzetti15a}).  The clusters were identified in three HST WFC3 pointings (field of view 2\farcm7 $\times$ 2\farcm7).  Two pointings span the radius of the inner disk of NGC\,7793, and include its nucleus, while the third pointing is in NGC\,628, just inside of the eastern inner disk edge.  Both NGC\,628 and NGC\,7793 are well-studied late-type spiral galaxies, and are relatively isolated relative to other massive systems.  However, they provide complementary views of star cluster populations.  Whereas NGC\,7793 is in the nearby Sculptor group at 3.6 Mpc, and is inclined, NGC\,628 is nearly face-on and is more distant, at 9 Mpc \citep{tully09a}.  Their star formation rates and stellar masses also differ by a factor of $\sim3$, and are 0.7 $M_{\odot}$ yr$^{-1}$ and $4 \times 10^9$ $M_{\odot}$ for NGC\,7793, and 2.0 $M_{\odot}$ yr$^{-1}$ and $1.6 \times 10^{10}$ $M_{\odot}$ for NGC\,628 \citep{lee09a, cook14a}. Thus NGC\,628 and 7793 represent examples towards the extremes of the full LEGUS sample of late-type spirals in terms of distance, orientation, mass, and star formation rate. This makes them a useful testbed for the performance of our analysis, one that will be extended to the full LEGUS sample in future work.

We have two goals in this paper. First, we wish to understand the performance of the stochastic stellar population analysis, including how it compares to traditional fitting methods, and to understand any systematic errors that might appear in the results due to choices of stellar tracks, stellar IMF, extinction law, assumed priors, or similar issues. Analysis of this sort has previously been performed by \citet{popescu12a} for the Large Magellanic Cloud, by \citet{fouesneau12a, fouesneau14a} for M31, and by \citet{de-meulenaer13a, de-meulenaer14a, de-meulenaer15a} for M31 and M33, but studies of systematics have mostly been limited to questions of isolated metallicity, evolutionary tracks, and extinction law. To our knowledge no previous work has covered the possible systematics thoroughly in the context of a single framework, and none at all have explored the issue of priors. Second, we wish to describe a portion of the LEGUS data pipeline, which will be used to produce star cluster catalogs based on \cs~modeling for the entire LEGUS sample. While this paper focuses on three fields in two galaxies for early analysis, the methods we develop here will be applied to the entire LEGUS sample. Ultimately, we will provide a catalog of homogeneously-observed and -analyzed star clusters, whose properties have been derived with a fully stochastic treatment of stellar population synthesis, for 50 galaxies that sample across the star-forming part of the Hubble sequence. This data set will provide tremendous insight into the extent to which star cluster formation and evolution depends on galactic environment.

The plan for the remainder of this paper is as follows. In \autoref{sec:methods} we describe the method we use to derive the physical properties of clusters, both stochastically and, for comparison, using a deterministic method. \autoref{sec:results} presents the results of this analysis for individual clusters, and discusses general trends and possible systematics. In this section we also investigate how the stochastic and deterministic analyses compare. \autoref{sec:popresults} extends this analysis to the properties of the star cluster population as a whole. Finally,  \autoref{sec:conclusions} summarizes our conclusions.

\section{Methodology}
\label{sec:methods}

\subsection{The Input Photometric Catalog}
\label{ssec:catalog}

A description of the steps required to produce final cluster catalogues of the LEGUS targets can be found in \citet{calzetti15a}, and in Adamo et al.~(in prep) we will present the custom pipelines we have developed to analyze the whole sample in a homogeneous fashion. We summarize here the main aspects of this process which led to the final cluster catalogues of the three pointings in two galaxies used in this work, i.e., NGC\,628 East (hereafter, NGC\,628e), NGC\,7793 West (NGC\,7793w), and NGC\,7793 East (NGC\,7793e). We create our catalogs by using the SExtractor algorithm \citep{bertin96a} on white-light images produced with the five standard HST LEGUS $UV-UBVI$ bands (WFC3 F275W and F336W, WFC3 or ACS F438W or F435W, F555W, and F814W, respectively). The exact filters and exposure times used in each pointing are listed in \autoref{tab:filters}. The cluster candidate catalogues contain only sources which satisfy the following criteria: 1) the V band concentration index (CI) must be greater than the stellar CI peak (the CI reference value slightly changes as function of galactic distance and HST camera); 2) the cluster candidate must be detected in two contiguous bands with a signal-to-noise higher than 3.

\capstartfalse
\begin{deluxetable*}{l@{\quad}c@{}c@{\quad}c@{}c@{\quad}c@{}c@{\quad}c@{}c@{\quad}c@{}c}
\tablecaption{List of filters\label{tab:filters} and exposure times}
\tablehead{
Field & Filter & $t_{\mathrm{exp}}$ (s) & Filter & $t_{\mathrm{exp}}$ (s) & Filter & $t_{\mathrm{exp}}$ (s) & Filter & $t_{\mathrm{exp}}$ (s) & Filter & $t_{\mathrm{exp}}$ (s)
}
\startdata
NGC\,628e & WFC3 F275W &2361 & WFC3 F336W &1119 &  ACS F435W &4720 &  WFC3 F555W & \phn965 & ACS 814W &1560  \\
NGC\,7793w & WFC3 F275W &2349 & WFC3 F336W &1101 & WFC3 F438W & \phn947 & WFC3 F555W & \phn680 & WFC3 F814W &\phn430 \\
NGC\,7793e & WFC3 F275W &2349 & WFC3 F336W &1101 & WFC3 F438W &\phn947 &  ACS F555W &1125  & ACS F814W &\phn971
\enddata
\end{deluxetable*}
\capstarttrue

The multi-band photometry of the cluster candidates of the three galactic fields used in this work has been derived with a fixed photometric aperture. We used an aperture radius of 4 and 5 px for NGC\,628 and NGC\,7793, respectively. The aperture radius is determined to include at least 50\% of the flux of a median built growth curve of several isolated clusters selected in the field of  each galaxy. The value will change mainly as function of galactic distance. The sky annulus 1 pixel wide is located at 7 pixel from the centre of the source. Averaged aperture corrections in all the filters have been estimated using manually selected isolated clusters in each galactic field. To take into account uncertainties produced by using a fixed aperture correction, we have added in quadrature the standard deviations of the aperture correction and the raw photometric error.

To remove contaminants from these automatically created catalogues we have visually inspected a subsample of cluster candidates in each galactic field with detections in at least 4 filters and absolute magnitudes brighter than M$_V=-6$ mag. Each inspected source has then been assigned a morphology quality flag. Class 1 objects are compact and centrally concentrated clusters; class 2 systems are clusters which show slightly elongated density profiles; class 3 are systems which show asymmetric profiles and multiple peaks, suggesting an association of stars; class 4 objects are single stars or artifacts, or any other spurious detection which can be excluded from the cluster catalogue. In this work we will only use high-confidence, visually-confirmed cluster catalogues clusters with photometry in at least 4 filters and flagged as class 1, 2, and 3. However, as a service for readers who might wish to make their own classifications and produce their own samples, in the electronic edition of the paper we provide a separate machine-readable table containing the results of a \cs~analysis of the clusters we omit here. We also note that, given their morphology, one might question whether class 3 objects should be called ``clusters". For the purposes of this analysis we treat them as such, but the fact that we have included them should be kept in mind when examining our results on the cluster population. Excluding them would produce a population that includes somewhat fewer young objects, since we find slightly younger than average ages for class 3 objects. However, since the focus of this work is on the method of analysis rather than the details of the population, we defer further discussion of this topic to future work.

Before proceeding, we make two additional notes regarding the catalog. First, we exclude the nuclear star cluster of NGC\,7793 from all the analysis presented below. Although this cluster is flagged to be in the high-confidence catalog, it is almost certainly not a simple stellar population, and we should not analyze its properties as if it were. Second, due to partial overlap of the NGC\,7793e and NGC\,7793w pointings, some clusters appear twice in the catalog, once for each field. We have retained the two separate entries in the analysis below, since they are observed with slightly different filters and thus do not produce completely identical results. However, when presenting population statistics in Section \ref{sec:popresults}, we include only the version of the cluster that appears in our NGC\,7793e catalog.

The final, high-confidence cluster catalog we use for this paper contains 621 distinct clusters. Counting the duplicates that appear in the overlapping fields for NGC\,7793 separately, the total rises to 645.

\subsection{Stochastic Models with \cs}

Our stochastic analysis makes use of the \cs~software package that is part of the \slug~(Stochastically Lighting Up Galaxies) software suite \citep{da-silva12a, da-silva14b, krumholz15b}. The Bayesian analysis performed by \cs~takes as input a set of absolute magnitudes $M_F$ in some set of filters, with corresponding errors $\Delta M_F$. It returns as output the posterior probability distribution for the cluster mass $M$, age $T$, and extinction $A_V$. Details of how this operation is performed can be found in the papers referenced above, and both the \slug~suite and the full software pipeline used in this paper are available at \url{https://www.slugsps.com}. Here we limit our discussion to the choice of two key inputs for the analysis: the libraries of simulated star clusters on which \cs~operates and the choice of prior probabilities that enter any Bayesian analysis. For a third input, the kernel density estimation bandwidth, we use $h=0.1$ dex in the physical variables and $h=0.1$ mag in the photometric ones.

\begin{figure}
\epsscale{1.2}
\plotone{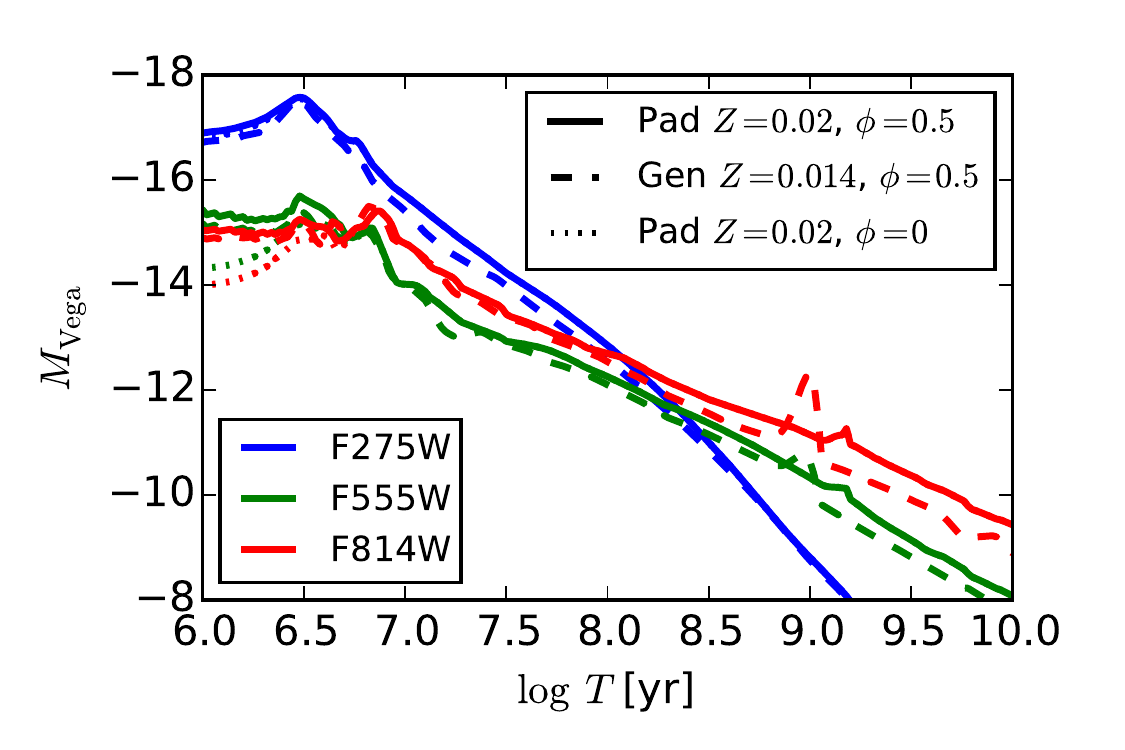}
\epsscale{1.0}
\caption{
\label{fig:trackcomp}
\red{Absolute Vega magnitude versus cluster age for Padova tracks with $Z=0.02$ and $\phi=0.5$, Geneva tracks with $Z=0.014$ and $\phi=0.5$, and Padova tracks with $Z=0.02$ and $\phi=0$, in the filters WFC3 F275W, WFC3 F555W, and WFC3 F814W, as predicted by \slug~for a $10^6$ $M_\odot$ cluster with a fully-sampled Kroupa IMF.
}
\\
}
\end{figure}

The \cs~code requires a library of simulated star clusters to act as a ``training set". We produce these using the \slug~stochastic stellar population synthesis code. To generate a library, we must specify the stellar evolution tracks, metallicity, extinction curve, stellar initial mass function (IMF), and the fraction of the ionizing light that is reprocessed into nebular emission within the observational aperture. One of the goals of this paper is to study how these choices affect the deduced cluster properties, and for this reason we generate a range of libraries, summarized in \autoref{tab:cslib}. The parameters we vary in these libraries are as follows: for tracks we consider both those from the Padova group including TP-AGB stars \citep{vassiliadis93a, girardi00a, vazquez05a} and from the Geneva group \citep{ekstrom12a}. For metallicities we consider $Z=0.004, 0.008, 0.020$ for the Padova tracks, and $Z=0.014$ for the Geneva tracks.\footnote{In principle it would be best to treat metallicity as a continuous variable to be fit, as we will fit cluster mass and age. \red{However, tracks including TP-AGB stars that are sufficiently finely sampled in metallicity to make such a treatment possible did not become available from the Padova group until late 2014, and have not yet been incorporated into \slug. No correspondingly fine tracks are available from the Geneva group yet. As a result, we will treat metallicity as fixed. As discussed below, this is not a significant limitation for our target galaxies, because the expected metallicity variation within the galaxy is relatively small.}} We consider extinction curves appropriate to both the Milky Way \citep{landini84a, fitzpatrick99a} and to a starburst \citep{calzetti00a}, and both \citet{kroupa01a} and \citet{chabrier05a} IMFs. Finally, for nebular emission we consider both $\phi=0.5$ (50\% of ionizing photons available to produce nebular emission inside the aperture) and $\phi=0$ (no nebular emission). \red{\autoref{fig:trackcomp} provides examples of how some of these choices affect the predicted variation of magnitude versus age in select filters. All the examples shown are for a $10^6$ $M_\odot$ cluster with a fully-sampled Kroupa IMF, and no extinction. One can clearly see in the \autoref{fig:trackcomp} the differences in predicted luminosity due to the treatment of TP-AGB stars at older ages, and due to the treatment of nebular emission at younger ages.}

\capstartfalse
\begin{deluxetable*}{lccccccccc}
\tablecaption{Model libraries for \cs\label{tab:cslib}}
\tablehead{
\colhead{Name} & 
\colhead{Tracks} & 
\colhead{IMF} &
\colhead{$Z$} &
\colhead{Extinction\tablenotemark{a}} &
\colhead{$\phi$\tablenotemark{b}} &
\colhead{$\log M$} &
\colhead{$\log T$} &
\colhead{$A_V$} &
\colhead{\# Realizations} \\
\colhead{} &
\colhead{} &
\colhead{} &
\colhead{} &
\colhead{} &
\colhead{} &
\colhead{($M_\odot$)} &
\colhead{(yr)} &
\colhead{(mag)} &
\colhead{}
}
\startdata
pad\_020\_kroupa\_MW\tablenotemark{c} & Padova & Kroupa & 0.020 & MW & 0.5 & $2-8$ & $5-10.18$ & $0-3$ & $10^7$ \\
pad\_020\_kroupa\_SB & Padova & Kroupa & 0.020 & SB & 0.5 & $2-8$ & $5-10.18$ & $0-3$ & $10^7$ \\
pad\_004\_kroupa\_MW & Padova & Kroupa & 0.004 & MW & 0.5 & $2-8$ & $5-10.18$ & $0-3$ & $10^7$ \\
pad\_008\_kroupa\_MW & Padova & Kroupa & 0.008 & MW & 0.5 & $2-8$ & $5-10.18$ & $0-3$ & $10^7$ \\
pad\_020\_chabrier\_MW & Padova & Chabrier & 0.020 & MW & 0.5 & $2-8$ & $5-10.18$ & $0-3$ & $10^7$ \\
gen\_014\_kroupa\_MW & Geneva & Kroupa & 0.014 & MW & 0.5 & $2-8$ & $5-\phn9.00$ & $0-3$ & $10^7$ \\
pad\_020\_kroupa\_MW\_noneb & Padova & Kroupa & 0.020 & MW & 0.0 & $2-8$ & $5-10.18$ & $0-3$ & $10^7$ 
\enddata
\tablenotetext{a}{MW = Milky Way extinction curve, SB = starburst extinction curve}
\tablenotetext{b}{$\phi$ is the fraction of the ionizing photons that produce nebular emission within the aperture; it combines the effects of a covering fraction $<1$ and some portion of the ionizing photons being absorbed directly by dust}
\tablenotetext{c}{Fiducial model\\}
\end{deluxetable*}
\capstarttrue

Our fiducial library, which we use for all purposes unless specified otherwise, is pad\_020\_kroupa\_MW. This uses Padova isochrones, metallicity $Z=0.02$, a Kroupa IMF, a Milky Way extinction curve, and 50\% of ionizing photons converted to nebular emission. We adopt this as our fiducial model because it appears to be the closest match to what we know of the target galaxies. Observed IMFs in local spiral galaxies are consistent with a universal function consistent with the \citet{kroupa01a} parameterization (see the recent reviews by \citet{krumholz14c} and \citet{offner14a}). Both NGC\,628 and NGC\,7793 have mean metallicities close to Solar, and for the typical $\sim 0.03$ dex kpc$^{-1}$ metallicity gradients observed in the inner disks of local spirals \citep[e.g.,][]{pilyugin04a}, the $<10$ kpc-wide regions spanned by our observed fields should have end-to-end metallicity differences of only a few tenths of a dex. On average we observe that, for the deterministic models at least (see below), clusters within face-on Solar metallicity spirals are better fitted by a Milky Way extinction curve \citep{cardelli89a} than with the alternatives. Finally, we use the Padova rather than Geneva evolutionary tracks because the majority of our clusters are older than $\sim 10$ Myr, and in this age range the superior treatment of AGB stars in the Padova models is advantageous. All these fiducial choices are discussed further in Adamo et al.~(2015, in preparation).

In addition to these choices, for each library we must specify how the masses, ages, and extinctions of the simple stellar populations are to be sampled. For all the libraries used in this paper, we choose a distribution that varies with cluster mass and age as
\begin{equation}
p_{\mathrm{lib}}(\vecx) = p_M(\log M) p_T(\log T) p_{A_V}(A_V)
\end{equation}
where
\begin{eqnarray}
p_M(\log M) & \propto &
\left\{
\begin{array}{l@{\quad}l}
1, & 2 < \log M < 4 \\
10^{-(\log M-4)}, & 4 \leq \log M < 8
\end{array}
\right.
\\
p_T(\log T) & \propto &
\left\{
\begin{array}{l@{\quad}l}
1, & 5 < \log T < 8 \\
10^{-(\log T - 8)}, & 8 < \log T < \log T_{\mathrm{max}}
\end{array}
\right. \\
p_{A_V} & \propto & 1, \quad 0 < A_V < 3
\end{eqnarray}
In the above expressions $M$ is in units of $M_\odot$, $T$ is in units of yr, and $A_V$ is in mag. We use $T_{\mathrm{max}} = 15$ Gyr for models using the Padova tracks, and $T_{\mathrm{max}} = 1$ Gyr for models using the Geneva tracks. This sampling distribution places most realizations at small masses and ages, where stochastic variation is largest, and uses fewer realizations at higher masses and older ages where they are smaller. We generate $10^7$ realizations for each library.

The final ingredient needed for a \cs~calculation is a prior probability distribution for the physical properties, which we use to weight the simulations in the library. One can demonstrate that the choice of priors will be important in at least some regimes via a simple thought experiment. Consider what might seem like a natural prior, a distribution that is flat in the log of the age. Stars' luminosities and colors do not change significantly until their ages reach $\sim 1$ Myr. Thus all star clusters younger than this age look identical, and the shape of the posterior probability distribution at ages below $\sim 1$ Myr must therefore be the same as the shape of the prior probability distribution. For a prior that is flat in log age, we would therefore conclude that age ranges of $\log (T/\mathrm{yr}) = 3.5-4$ and $\log (T/\mathrm{yr}) = 5.5-6$ are equally likely. This seems unlikely to be correct: even if it were completely unbound, a stellar population formed $<1$ Myr ago would still be close enough together to be recorded as a cluster in the LEGUS catalog, so every cluster that reaches an age of $\log (T/\mathrm{yr}) = 3.5-4$ must eventually reach an age of $\log (T/\mathrm{yr}) = 5.5-6$. However, this cluster will reside in the interval $\log (T/\mathrm{yr}) = 3.5-4$ for a mere 6,800 yr, while it will spend 680,000 yr in the interval $\log (T/\mathrm{yr}) = 5.5-6$. Clearly there should be 100 times as many clusters in the latter bin as in the former, making the latter 100 times as likely. Even worse, suppose further that we have a cluster whose colors admit ages older as well as younger than $\sim 1$ Myr. In this case the weight that we would end up assigning to the larger ages would depend strongly on whether our model grid contained a youngest age of, say, $10^3$ yr, $10^4$ yr, $10^5$ yr, or $10^6$ yr, because the amount of probability ``phase space" allowed at young ages would depend on this choice. Clearly some attention to priors is required to avoid outcomes such as this.

For ages younger than a few Myr, the proper prior is almost certainly flat in age. This is because, even if a newly-formed collection of stars has negligible gravitational binding energy, at the typical velocity dispersions of a few km s$^{-1}$ found in young clusters, it will disperse over a region no more than $\sim 10$ pc in size over this time. In an extragalactic survey such as LEGUS, it would still be detected as a cluster. Thus cluster dispersal is irrelevant at such young ages. Similarly, as noted above, stellar evolution effects are also negligible. Given the absence of any physical mechanism to bias the age distribution, all ages are equally likely, suggesting that the proper prior is flat in age.

For ages greater than $\sim 10^{6.5}$ yr, the proper prior is significantly less certain, as there is an extensive debate in the literature on this topic. Some authors report distributions $dN/dT \propto T^{-1}$ \citep[i.e., flat in log age; e.g.][]{fall05a, fall09a, chandar10a, chandar10b, chandar11a, fall12a, fouesneau12a, popescu12a}, while others (mostly focusing on ages $>10$ Myr) find $dN/dT \propto T^0$ \citep[i.e., flat in age; e.g.][]{boutloukos03a, de-grijs03b, gieles07a, bastian11a, bastian12a, bastian12b, silva-villa14a}; some authors also report intermediate results, with the index of the age distribution changing as a function of age \citep[e.g.][]{fouesneau14a}. The results appear to depend in part on the criteria one uses to define what is a cluster and thus should be included in the cluster catalog; see \citet{krumholz14c} for a recent review. The situation is further complicated by the fact that the prior we want to use is not the intrinsic distribution of star cluster ages, but the convolution of that with our selection function. Properly modeling this would require much greater understanding of our observational completeness and biases than we now possess. Given these complexities, as a fiducial prior we choose a compromise, $dN/dT \sim T^{-0.5}$, which is intermediate between flat in age ($dN/dT \sim\mathrm{constant}$) and flat in log age ($dN/d\log T\sim\mathrm{constant}$, so that $dN/dT\sim T^{-1}$). However, we will explore the effects of varying this choice below. The maximum possible age will be set by the maximum age in our library.

For the prior on the mass distribution, we note that observations of young star clusters consistently find $dN/dM\propto M^{-2}$ \citep[e.g.,][]{williams97a, zhang99a, larsen02a, bik03a, de-grijs03a, goddard10a, bastian11a, fall12a, fouesneau12a}, with relatively little variation. These results are mostly derived at masses high enough that stochasticity is unimportant, and thus it seems reasonable simply to extrapolate them into our regime. We therefore adopt $p(\log M) \propto M^{-1}$ (corresponding to $dN/dM \propto M^{-2}$) as our fiducial prior on the mass distribution. Two caveats are worth mentioning, though. First, as with the age distribution, the truly correct prior to use would be the convolution of this function with our observational selection, which unfortunately is not fully characterized as yet. Second, to avoid a logarithmic divergence, we must truncate the prior distribution at both the high and low mass ends; we implicitly choose the truncation values via the range of masses we include in our library. Fortunately the choices here have relatively little effect, because our library covers masses from $10^2 - 10^8$ $M_\odot$, which spans the entire plausible mass range: the lower limit of $100$ $M_\odot$ is almost certainly smaller than the smallest cluster we have any hope of detecting, and, given the size of our sample, the upper limit is much more massive than the most massive cluster we would expect to find even if the underlying cluster mass function continued as $dN/dM \propto M^{-2}$ to even higher masses.

Finally, for the $A_V$ distribution, observations indicate that $A_V$ values tend to be at most $\sim 0.5$ at ages above $\sim 10$ Myr \citep[e.g.,][]{whitmore02a, whitmore11a, mengel05a, bastian14a, hollyhead15a}. At younger ages,  \citet{fouesneau12a, fouesneau14a} report a fairly broad distribution of $A_V$. In principle we could use a prior that combines age and $A_V$ to disfavor a combination of old age and high-$A_V$. However, the functional form of this constraint is not well-characterized, and we wish to keep our priors on age and $A_V$ independent to ease the analysis of how our results depend on a choice of priors. For this reason we choose to adopt a flat distribution in $A_V$ as a prior. 

In summary, our fiducial prior distribution follows $p_{\mathrm{prior}}(\vecx) \propto M^{-1} T^{-0.5}$, with no dependence on $A_V$. However, in \autoref{ssec:priors} we will check the dependence of our results upon this choice.

\subsection{Deterministic Models and Traditional $\chi^2$-Fitting Method}

The standard photometric analysis of the LEGUS cluster catalogues is performed with \yg\footnote{\url{http://ttt.astro.su.se/projects/yggdrasil/yggdrasil.html}} \citep{zackrisson11a} deterministic models and a traditional $\chi^2$-fitting procedure \citep[e.g., see][]{calzetti15a}. \yg~is a spectral synthesis code that can provide spectral and integrated flux tables for a large set of stellar and nebular parameters. In \yg, the spectrum produced by the stellar population at each age step is then used as input for a calculation of the nebular emission using \texttt{cloudy} \citep{ferland13a} to ensure a self-consistent treatment.

For the LEGUS cluster analysis in this paper, we have carefully matched the physical models used in \yg~to those of the fiducial \cs~library. Thus for our \yg~models we assume that clusters are a simple stellar population, and we compute the stellar spectrum using \texttt{Starburst99} \citep{leitherer99a, vazquez05a} atmosphere models. We use the \citet{kroupa01a} IMF, the same Padova AGB stellar tracks used in the \cs~models, and a metallicity $Z=0.02$. We calculate nebular emission assuming that (1) the metallicity of gas and stars is the same, (2) the average gas density is 100 cm$^{-3}$, (3) the ionizing flux is normalized by setting it to that of a $10^6$ $M_\odot$ stellar population, and (4) the covering factor set to 50\%, meaning that 50\% of the ionizing photons are assumed to be absorbed by hydrogen atoms within the observational aperture. To model extinction, we adopt a Milky Way extinction curve, and consider differential extinctions E(B$-$V) between 0 and 1.5 mag with steps of 0.01 mag. We apply extinction to the model spectra before convolving them with the filter throughput. Thus in every respect except nebular emission, which we discuss in \autoref{sssec:caveat}, our \yg~models are completely matched to our fiducial \cs~one.

The $\chi^2$-fitting algorithm used to derive cluster physical properties has been presented and tested in \citet{adamo10a}. The algorithm compares the observed photometry to the library of \yg~models, and returns the library cluster age, extinction, and mass that deviates least from the observations. A quality of the fit value is estimated from the reduced $\chi^2$, i.e., $Q$-parameter. Good fits have $Q$ values close to 1, while values below 0.1 suggest poor fits to the observed cluster SEDs. The $\chi^2$-fitting algorithm has also been extended by \citet{adamo12a} to produce uncertainties in the derived cluster physical properties contained within the 68\% confidence limits of the solution parameter space \citep[their Figure 9]{calzetti15a}.  

\subsubsection{A Caveat Regarding Nebular Emission}
\label{sssec:caveat}

Before presenting results, we pause to add a caveat to our comparison of the results obtained with \yg~and \cs. We have tried to ensure that the underlying physical models are as similar as possible in the two codes, so that any differences between them arise purely from the fact that \cs~includes stochasticity and returns a full posterior PDF, while our models using \yg~are non-stochastic, and our fits to them use a traditional $\chi^2$ method. The one quantity we have not been able to match precisely is the treatment of nebular emission.

To good approximation, an H~\textsc{ii} region produces a nebular luminosity per ionizing photon injected that is a function of the gas metallicity and ionization parameter. While the \yg~and \cs~models are matched in metallicity, they are not precisely matched in ionization parameter. As noted above, the nebular contribution in the \yg~models we use in this paper has been derived by taking spectra computed for a $10^6$ $M_\odot$ stellar population illuminating an H~\textsc{ii} region with a constant density of $10^2$ cm$^{-3}$ and a 50\% covering fraction. (See \citet{zackrisson11a} for further details.) In the \yg~models, the inner radius of the H~\textsc{ii} region scales with total luminosity to the 1/2 power, so the ionization parameter at the illuminated face of the H~\textsc{ii} region is kept roughly constant -- roughly rather than precisely because the ratio of ionizing to total luminosity is not constant as a stellar population ages. However, because the volume of ionized gas changes as the luminosity does, and the ionizing photon flux changes as the photons propagate through the nebula, this prescription causes the volume-averaged ionization parameter to vary significantly with the mass and age of the stellar population. The mean is close to $\log\,\mathcal{U} = -2.5$, but there is a non-negligible spread.

\texttt{Slug}'s treatment of nebular emission also assumes constant density within an H~\textsc{ii} region, and we have chosen the same 50\% covering fraction as in the \yg~models. However, in its simplest form, \slug~assumes that the volume-averaged ionization parameter has a constant value $\log\,\mathcal{U} = -3$, while the ionization parameter at the inner edge is not held constant.\footnote{Full details of the \slug~method, and the computational motivations for choosing a constant ionization parameter, are given in \citet{krumholz15b}.} While \slug~also supports an option to pass the spectra through \texttt{cloudy}, and thus in principle we could fully emulate \yg's assumptions, it is not computationally feasible to run \texttt{cloudy} on all, or even a substantial subset, of the nearly $10^8$ models in our libraries. Thus we are left with an imperfect match in ionization parameter between the \cs~and \yg~models; the \yg~models assume that H~\textsc{ii} regions form a sequence where the ionization parameter at the illuminated face is fixed but the volume-averaged ionization parameter is not, while the \cs~models fix the volume-averaged but not the illuminated face ionization parameters.

How does this difference in ionization parameter translate into a difference in photometry, which is what we ultimately care about? Obviously the difference will be negligible in stellar populations older than $\sim 4$ Myr, when the ionizing luminosity drops precipitously. Even in younger populations, the nebular contribution to the total luminosity is only $\sim 10\%$ for F275W and F438W filters, and thus a difference in how this emission is treated is again unimportant. The nebular contribution is most significant for F336W, F555W, and F814W, where it can reach $\sim 60-70\%$ of the total \citep[e.g.,][]{reines10a}.

In the case of F555W, the nebular contribution is dominated by the [O~\textsc{iii}] $\lambda\lambda 4959, 5007$ and H$\beta$ lines, with additional contributions from bound-free and two-photon continuum emission, and from the H$\alpha$ and [N~\textsc{ii}] $\lambda6584$ lines (which are intrinsically bright but lie near the edge of the filter response curve, and will be shifted out of the filter entirely for even modest redshifts). Over the observationally-plausible range of ionization parameters (roughly $\log\,\mathcal{U} = -3$ to $-2$ -- e.g., see the compilation in \citealt{verdolini13a}), both Balmer line and bound-free production per ionizing photon injected vary by a factor of $\sim 2$. The [O~\textsc{iii}] emission, as well as the other metal lines, are sensitive to both the temperature and the ionization state of the nebula. Consequently, they can easily vary by an order of magnitude as the ionization parameter changes \citep[e.g.,][]{kewley01a, yeh13a, verdolini13a}, and at the highest ionization parameters they can be a factor of several brighter than the H$\beta$ line. For F814W, the nebular contribution is dominated by free-free and bound-free continuum, with an a sub-dominant contribution of Paschen lines and [S~\textsc{iii}] $\lambda\lambda9069,9532$. The two continuum sources and the Paschen lines are relatively insensitive to the ionization parameter, varying by less than a factor of 2. In contrast, the [S~\textsc{iii}] lines, while they are subdominant, have the largest potential variation. Their strength can change by at least an order of magnitude over the plausible range of ionization parameters. Finally, the nebular contribution F336W is strongly dominated by the bound-free and two-photon emission. The former varies by less than a factor of 2 over the plausible ionization parameter range, while the latter stays almost entirely constant until the density reaches $\sim 10^4$ cm$^{-3}$, at which point the two-photon process is collisionally quenched.

Taken together, these factors suggest that differences in the assumed ionization parameter between \yg~and \cs~will lead to at most factor of $\sim 3$ differences in F336W, F555W, and F814W luminosity in young stellar populations. These variations are not trivial, but we shall see below that they constitute a fairly minor uncertainty in comparison to those associated with degeneracies between age and extinction, or variations due to stochasticity. However, as a final caution, we note that the prescriptions for nebular emission in both \cs~and \yg~are only rough approximations to the real complexity of H~\textsc{ii} regions. For example, both \cs~and \yg~assume a fixed, age-independent covering fraction. However, in reality H~\textsc{ii} regions expand with time while the observational aperture remains fixed, so the covering fraction should drop with age \citep[cf.][]{whitmore11a}, at a rate that depends on both the cluster's luminosity and the density of its surrounding, both of which affect the rate at which its H~\textsc{ii} regions expands. Thus age estimates below $\sim 5$ Myr should be regarded with extreme caution regardless of the fitting method.

\section{Results for Individual Clusters}
\label{sec:results}

Throughout this section we will refer to the results of an analysis of the cluster populations of NGC\,7793e, NGC\,7793w, and NGC\,628e performed using \cs. For the convenience of other authors who wish to use our results without having to rerun the full analysis, we summarize the \cs~output in \autoref{tab:summary}. A full machine-readable version of this Table is included in the electronic edition of this paper, along with an analogous table containing the same results for the objects we have classified as unlikely to be genuine star clusters. We provide the latter as a service for those who might wish to make their own classifications.

\capstartfalse
\begin{turnpage}
\begin{deluxetable*}{lccccccccccccccccccccccc}
\tabletypesize{\scriptsize}
\tablecaption{Summary marginal posterior PDFs\label{tab:summary}}
\tablehead{
\colhead{ID} & \colhead{Mode class\tablenotemark{a}} & \colhead{Mean class\tablenotemark{a}} & \colhead{} & 
\multicolumn{5}{c}{$\log (M/M_\odot)$ percentiles} & \colhead{} &
\multicolumn{5}{c}{$\log (T/\mathrm{yr})$ percentiles} & \colhead{} &
\multicolumn{5}{c}{$A_V$ percentiles} & \colhead{} & 
\colhead{$D_5$ [mag]} & \colhead{$D_5^{\mathrm{norm}}$} \\[0.5ex]
\cline{5-9} \cline{11-15} \cline{17-21}\\[-0.5ex]
\colhead{} & \colhead{} & \colhead{} & \colhead{} &
\colhead{16} & \colhead{25} & \colhead{50} & \colhead{75} & \colhead{84} & \colhead{} &
\colhead{16} & \colhead{25} & \colhead{50} & \colhead{75} & \colhead{84} & \colhead{} &
\colhead{16} & \colhead{25} & \colhead{50} & \colhead{75} & \colhead{84} & \colhead{} &
\colhead{} & \colhead{} 
}
\startdata
\cutinhead{NGC\,628}
\sidehead{pad\_020\_kroupa\_MW, $\beta=-2.0$, $\gamma=-0.0$}
3 & 1.00 & 1.00 & & 4.33 & 4.38 & 4.48 & 4.58 & 4.63 & &8.10 & 8.18 & 8.30 & 8.42 & 8.46 & &0.09 & 0.14 & 0.26 & 0.43 & 0.52 & &0.03 & \phn0.51 \\
4 & 1.00 & 1.00 & & 3.79 & 3.84 & 3.99 & 4.09 & 4.19 & &6.14 & 6.18 & 6.34 & 6.51 & 6.71 & &0.61 & 0.76 & 0.97 & 1.13 & 1.20 & &0.08 & \phn1.69 \\
5 & 2.00 & 1.67 & & 2.50 & 2.65 & 3.00 & 3.34 & 3.49 & &6.22 & 6.30 & 6.43 & 6.59 & 6.71 & &0.71 & 0.80 & 0.97 & 1.11 & 1.18 & &0.05 & \phn0.89 \\
\sidehead{pad\_020\_kroupa\_MW, $\beta=-2.0$, $\gamma=-0.5$}
3 & 1.00 & 1.00 & & 3.44 & 4.29 & 4.43 & 4.53 & 4.58 & &7.00 & 8.02 & 8.26 & 8.38 & 8.42 & &0.12 & 0.17 & 0.33 & 0.59 & 0.87 & &0.03 & \phn0.51 \\
4 & 1.00 & 1.00 & & 3.79 & 3.84 & 3.94 & 4.09 & 4.14 & &6.10 & 6.18 & 6.30 & 6.47 & 6.55 & &0.71 & 0.80 & 0.99 & 1.16 & 1.23 & &0.08 & \phn1.69 \\
5 & 2.00 & 1.67 & & 2.50 & 2.65 & 2.95 & 3.25 & 3.39 & &6.22 & 6.30 & 6.43 & 6.55 & 6.63 & &0.73 & 0.80 & 0.97 & 1.11 & 1.18 & &0.05 & \phn0.89 \\
\sidehead{pad\_020\_kroupa\_MW, $\beta=-2.0$, $\gamma=-1.0$}
3 & 1.00 & 1.00 & & 2.85 & 3.05 & 3.54 & 4.48 & 4.53 & &6.75 & 6.79 & 7.04 & 8.26 & 8.34 & &0.19 & 0.31 & 0.78 & 1.11 & 1.23 & &0.03 & \phn0.51 \\
4 & 1.00 & 1.00 & & 3.79 & 3.84 & 3.94 & 4.04 & 4.14 & &6.10 & 6.18 & 6.30 & 6.43 & 6.51 & &0.76 & 0.83 & 0.99 & 1.16 & 1.23 & &0.08 & \phn1.69 \\
5 & 2.00 & 1.67 & & 2.45 & 2.60 & 2.95 & 3.20 & 3.34 & &6.22 & 6.26 & 6.39 & 6.51 & 6.59 & &0.73 & 0.80 & 0.97 & 1.11 & 1.18 & &0.05 & \phn0.89 \\
\sidehead{pad\_020\_kroupa\_SB, $\beta=-2.0$, $\gamma=-0.0$}
3 & 1.00 & 1.00 & & 4.34 & 4.39 & 4.49 & 4.59 & 4.64 & &8.10 & 8.18 & 8.30 & 8.38 & 8.46 & &0.12 & 0.17 & 0.31 & 0.50 & 0.61 & &0.04 & \phn0.50 \\
4 & 1.00 & 1.00 & & 3.85 & 3.90 & 4.05 & 4.19 & 4.24 & &6.14 & 6.22 & 6.34 & 6.55 & 6.83 & &0.57 & 0.83 & 1.11 & 1.32 & 1.39 & &0.09 & \phn1.90 \\
5 & 2.00 & 1.67 & & 2.61 & 2.81 & 3.11 & 3.40 & 3.55 & &6.22 & 6.30 & 6.43 & 6.63 & 6.79 & &0.71 & 0.85 & 1.09 & 1.28 & 1.35 & &0.08 & \phn1.28 \\
\cutinhead{NGC\,7793e}
\sidehead{pad\_020\_kroupa\_MW, $\beta=-2.0$, $\gamma=-0.0$}
5 & 2.00 & 2.00 & & 3.34 & 3.44 & 3.59 & 3.74 & 3.79 & &7.65 & 7.73 & 7.89 & 8.06 & 8.14 & &0.14 & 0.19 & 0.38 & 0.61 & 0.73 & &0.06 & \phn0.74 \\
8 & 2.00 & 1.67 & & 3.39 & 3.49 & 3.64 & 3.74 & 3.79 & &7.49 & 7.57 & 7.73 & 7.85 & 7.93 & &0.07 & 0.12 & 0.24 & 0.43 & 0.61 & &0.02 & \phn0.30 \\
10 & 3.00 & 3.00 & & 2.21 & 2.30 & 2.55 & 2.85 & 2.95 & &5.49 & 5.57 & 5.77 & 5.98 & 6.06 & &1.11 & 1.16 & 1.30 & 1.44 & 1.51 & &0.04 & \phn0.53 \\
\sidehead{pad\_020\_kroupa\_MW, $\beta=-2.0$, $\gamma=-0.5$}
5 & 2.00 & 2.00 & & 3.30 & 3.39 & 3.59 & 3.74 & 3.79 & &7.61 & 7.65 & 7.85 & 8.02 & 8.10 & &0.14 & 0.21 & 0.43 & 0.66 & 0.78 & &0.06 & \phn0.74 \\
8 & 2.00 & 1.67 & & 3.05 & 3.34 & 3.59 & 3.74 & 3.79 & &6.79 & 7.32 & 7.65 & 7.81 & 7.89 & &0.09 & 0.14 & 0.33 & 0.99 & 1.18 & &0.02 & \phn0.30 \\
10 & 3.00 & 3.00 & & 2.21 & 2.30 & 2.55 & 2.85 & 2.95 & &5.49 & 5.57 & 5.77 & 5.98 & 6.06 & &1.11 & 1.16 & 1.30 & 1.44 & 1.51 & &0.04 & \phn0.53 \\
\sidehead{pad\_020\_kroupa\_MW, $\beta=-2.0$, $\gamma=-1.0$}
5 & 2.00 & 2.00 & & 3.25 & 3.34 & 3.54 & 3.69 & 3.74 & &7.53 & 7.61 & 7.77 & 7.93 & 8.02 & &0.17 & 0.26 & 0.47 & 0.73 & 0.87 & &0.06 & \phn0.74 \\
8 & 2.00 & 1.67 & & 2.50 & 2.75 & 3.44 & 3.64 & 3.74 & &6.59 & 6.67 & 7.28 & 7.69 & 7.81 & &0.17 & 0.26 & 0.97 & 1.28 & 1.37 & &0.02 & \phn0.30 \\
10 & 3.00 & 3.00 & & 2.21 & 2.30 & 2.55 & 2.85 & 2.95 & &5.49 & 5.57 & 5.77 & 5.98 & 6.06 & &1.11 & 1.16 & 1.30 & 1.44 & 1.51 & &0.04 & \phn0.53 \\
\sidehead{pad\_020\_kroupa\_SB, $\beta=-2.0$, $\gamma=-0.0$}
5 & 2.00 & 2.00 & & 3.40 & 3.50 & 3.65 & 3.80 & 3.85 & &7.61 & 7.69 & 7.85 & 8.02 & 8.10 & &0.21 & 0.31 & 0.57 & 0.87 & 1.04 & &0.05 & \phn0.64 \\
8 & 2.00 & 1.67 & & 3.35 & 3.45 & 3.65 & 3.75 & 3.80 & &7.32 & 7.53 & 7.73 & 7.85 & 7.93 & &0.09 & 0.14 & 0.26 & 0.57 & 1.11 & &0.02 & \phn0.31 \\
10 & 3.00 & 3.00 & & 2.22 & 2.32 & 2.61 & 2.91 & 3.01 & &5.49 & 5.61 & 5.82 & 5.98 & 6.06 & &1.30 & 1.35 & 1.49 & 1.63 & 1.72 & &0.07 & \phn0.76 \\
\enddata
\tablenotetext{a}{Mode and mean of the classifications given by the visual classifiers; 0 = source was not visually classified (too faint); 1 = symmetric, compact cluster; 2 = concentrated object with some degree of asymmetry or color gradient; 3 = diffuse or multiple peak system, possibly spurious alignment; 4 = probable spurious detection (foreground/background source, single bright star, artifact}
\tablecomments{\autoref{tab:summary} appears in its entirety in the electronic edition of The Astrophysical Journal. A portion is shown here for guidance regarding its form and content. The full table includes 645 visually-confirmed clusters\red{, of which 621 are unique and 24 are overlapping between NGC\,7793e and NGC\,7793w. It also includes} 2326 objects visually-classified as unlikely to be clusters (provided in separate files). }
\end{deluxetable*}
\end{turnpage}

\capstarttrue

\subsection{Are the \cs~Libraries Consistent with the Observations?}
\label{ssec:consistency}

\begin{figure}
\epsscale{1.2}
\plotone{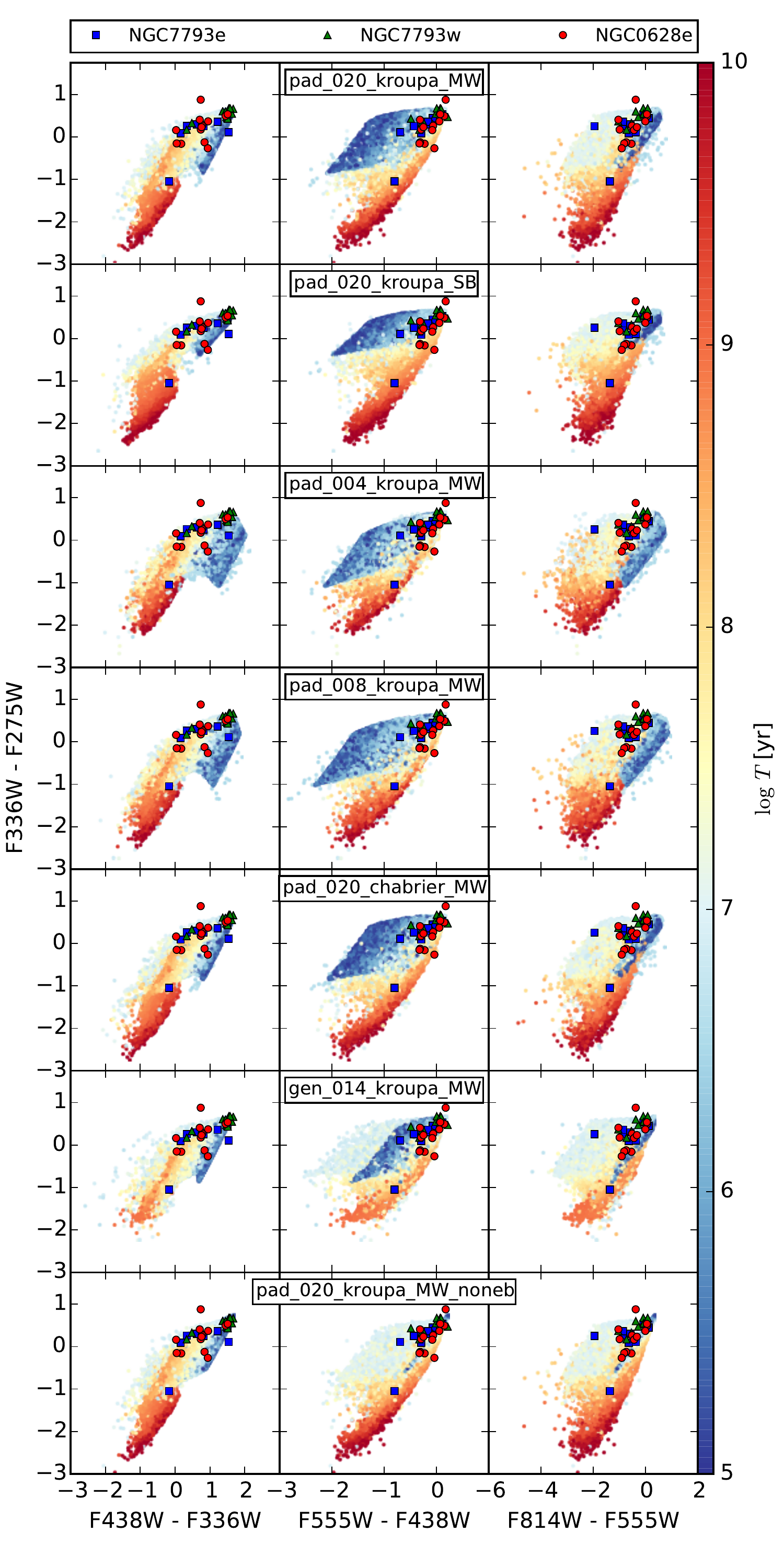}
\epsscale{1.0}
\caption{
\label{fig:color_color}
Color-color plots comparing our model libraries against the observed star clusters in NGC\,7793e, NGC\,7793w, and NGC\,628. In all panels the $y$ axis shows the color $\mathrm{F336W}-\mathrm{F275W}$, while the $x$ axis shows the color indicated. (Note that, for NGC\,628e we have used F435W in place of F438W -- see \autoref{tab:filters}.) Filled markers show every 20th cluster in the observed photometric catalogs, with the symbol shape and color indicating the galaxy as shown in the legend; the filters used for the photometric points are those indicated in \autoref{tab:filters}. Colored points show 1\% of the clusters from each of the \cs~libraries, selected at random; points shown for the \cs~libraries are all for WFC3 filters. Each row corresponds to one of the libraries listed in \autoref{tab:cslib}, as indicated in the right column. Points are colored by the age of the model, as indicated in the color bar.\\
}
\end{figure}

As a first, most basic question, we examine the extent to which the libraries of stochastic models reproduce the colors and magnitudes found in the observed sample. The kernel density estimation method used by \cs~will return results even if the correspondence between the model libraries and the observations is very poor, but those results should be considered reliable only to the extent that the models and observations are in reasonable correspondence. To check whether this is the case, in \autoref{fig:color_color} we show the distribution of our \cs~model libraries in color-color space, as compared to the observed catalog. As the plot shows, the model libraries generally cover loci in color-color space very similar to the observations.

By itself, agreement in cuts in color-color space does not confirm that our libraries are a good analog to the observations. Even if the real and synthetic data show similar distributions in particular projections, they may be differently distributed in higher dimensions. Moreover, unlike in deterministic models like \yg, the total mass and absolute magnitude are not free parameters in the \cs~models. Because the amount of stochasticity depends on how well the mass distribution is sampled, the dispersion in color at fixed age is a function of the absolute magnitude. As a result, it is possible that our models cover the same range as the data in color-color space, but that they might not in color-magnitude space.

To perform a more quantitative comparison of the observed and synthetic catalogs, we therefore examine the distribution of distances in photometric space between the observed star clusters and their closest analogs in our synthetic catalogs. We define the absolute and normalized photometric distances between an observed star cluster and the $i$th member of a \cs~library by
\begin{eqnarray}
\label{eq:photdist}
D & = & \sqrt{\frac{1}{N} \sum_j \left(M_{F_j,\mathrm{obs}} - M_{F_j,i}\right)^2}, \\
\label{eq:photdistnorm}
D^{\mathrm{norm}} & = & \sqrt{\frac{1}{N} \sum_j \left(\frac{M_{F_j,\mathrm{obs}} - M_{F_j,i}}{\Delta M_{F_j,\mathrm{obs}}}\right)^2}
\end{eqnarray}
where $N$ is the number of filters, $M_{F_j,\mathrm{obs}}$ is the magnitude of the observed cluster in filter $F_j$, $\Delta M_{F_j,\mathrm{obs}}$ is the observational error on this value, and $M_{F_j,i}$ is the magnitude of the $i$th library cluster in filter $F_j$. The exact filters used in this comparison are those given in \autoref{tab:filters}, so generally $N=5$; however, there are a very small number of clusters which lie outside the image in one of the filters, and for these $N=4$.

\begin{figure}
\plotone{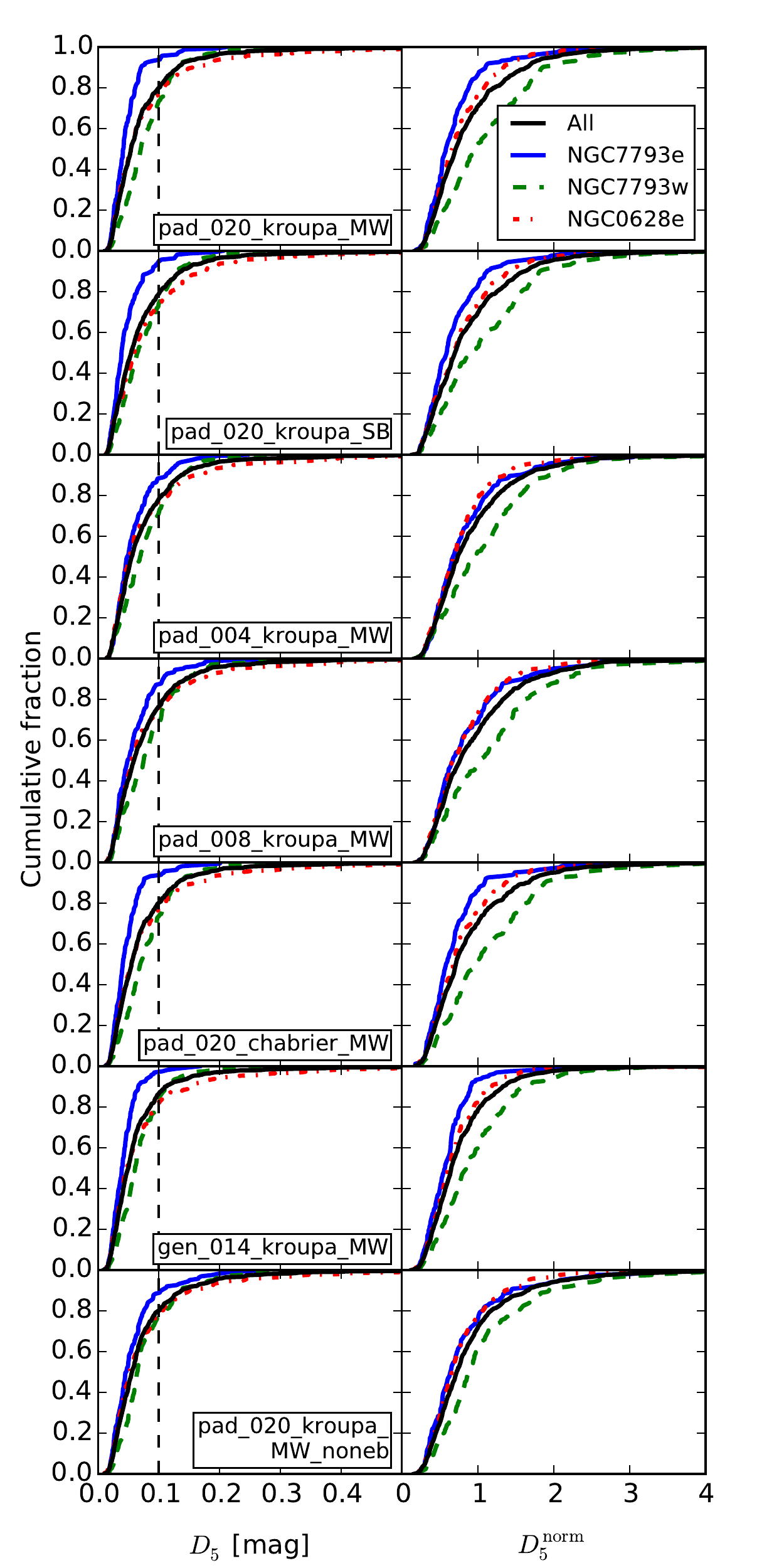}
\caption{
\label{fig:error_dist}
Cumulative distribution functions of photometric distances between the observed star clusters and the 5th nearest neighbor in our synthetic \cs~catalogs. The left column shows the raw 5th-nearest neighbor photometric distance $D_5$ (\autoref{eq:photdist}), while the right column shows $D_5^{\mathrm{norm}}$ (\autoref{eq:photdistnorm}), the 5th nearest neighbor distance normalized to the photometric errors. Each row is for a different \cs~library, as indicated by the labels in the left column, and different line colors and styles are for different galaxies, as indicated by the legend. The black line, marked ``All", is the summed CDF for the three fields. The vertical dashed lines in the left column indicate $D = 0.1$ mag, the bandwidth we use for kernel density estimation.\\
}
\end{figure}

For each observed cluster, we define $D_5$ and $D_5^{\mathrm{norm}}$ as the absolute and normalized distances to the 5th nearest neighbor in one of the \cs~libraries. (Note that the 5th nearest neighbor is not necessarily the same simulated cluster for the absolute and normalized distances.) \autoref{fig:error_dist} shows the cumulative distribution of $D_5$ and $D^{\mathrm{norm}}_5$ for each of our sample galaxies compared to each of our model catalogs; plots for the $n$th nearest neighbor, with $n \sim 1 - 20$, are qualitatively similar. Clearly nearly all of our observed star clusters have close analogs in our synthetic libraries. For our fiducial model library, pad\_020\_kroupa\_MW, we find that 71\% of observed clusters have at least 5 matches in our library within the $1\sigma$ photometric errors (i.e., 71\% have $D_5^{\mathrm{norm}} < 1$), 95\% have at least 5 matches within the $2\sigma$ photometric errors, and 99\% have 5 matches within the $3\sigma$ errors. The figures are comparable for the other model libraries, and the differences between them are quite minor. The normalized distances are generally smallest for NGC\,0628e and largest for NGC\,7793e, but this is more a reflection of the size of the photometric errors in those fields than of any intrinsic differences between the cluster populations.

While the nearness of our library points to the observations is encouraging, we should inquire a bit further about what the distribution of $n$th nearest neighbor distances \textit{should} look like if our models are in fact a good fit. For a Poisson process, the cumulative distribution function of $n$th nearest neighbor distances $r$ for a point at position $\vecx$ is given by
\begin{equation}
\label{eq:ndist}
d_n(r \mid \vecx) = 1 - \sum_{k=0}^{n-1} \frac{\lambda(r,\vecx)^k e^{-\lambda(r,\vecx)}}{k!},
\end{equation}
where $\lambda(r,\vecx)$ is the expected number of points in a ball of radius $r$ centered on $\vecx$. Intuitively, this is simply the statement that the distribution of $n$th nearest neighbor distances is 1 minus the probability that there are $n-1$ or fewer points within a ball of size $r$ around the point in question. Note that we are free to use any metric to measure $r$, so we can use the normalized photometric distance as well as the absolute one. Evaluating the expectation value $\lambda$ requires knowledge of the shape of the underlying probability distribution around the point of interest. In principle we could evaluate this for each of our points from the kernel density estimate for the PDF, but doing so would be quite computationally intensive, and for small distances would likely depend on our choice of bandwidth parameter. Instead, we can qualitatively check whether our nearest neighbor distribution is consistent with our models being a good match to the data by making a few simplifying assumptions that allow us to evaluate $\lambda$ analytically. 

\begin{figure}
\plotone{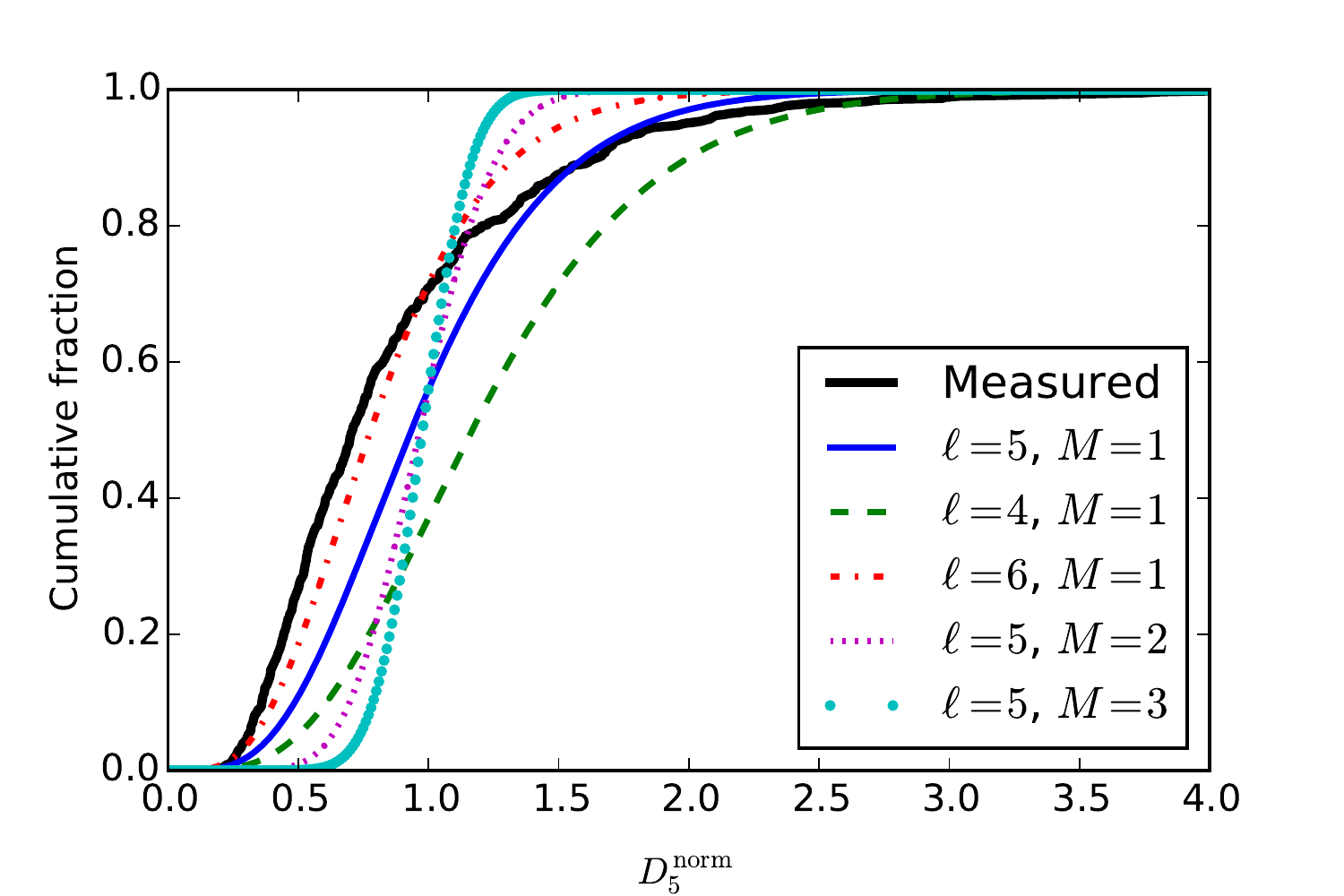}
\caption{
\label{fig:d5expectation}
Cumulative distribution function of 5th nearest neighbor photometric distances normalized by the observed photometric errors, $D_5^{\mathrm{norm}}$. The thick black line shows the distribution we measure summed over all our sample fields and using our fiducial model library, pad\_020\_kroupa\_MW. The various color lines show the expected distribution of 5th nearest neighbor distances, computed using equations (\ref{eq:ndist}) and (\ref{eq:lambdadist}) for various values of the expected number of library models within the photometric error circle, $\ell$, and the dimensionality of the data, $M$, as indicated in the legend.\\
}
\end{figure}

Suppose that our library is large enough that we expect there to be $\ell$ simulations from the library within a normalized photometric distance $D^{\mathrm{norm}} = 1$ of each observed point. Further suppose that the library of simulation points lies along an $M$-dimensional manifold within our 5-dimensional photometric space; for example, if the simulations near a particular point mostly lie along a line in photometric space, then the number within a given photometric distance increases linearly with distance, and $M=1$. If they are mostly along a plane then $M=2$, and so forth. In this case we have
\begin{equation}
\label{eq:lambdadist}
\lambda(r) = \ell r^M,
\end{equation}
and we can evaluate the expected $n$th nearest neighbor distribution $d_n(r)$ directly. \autoref{fig:d5expectation} shows how the measured distribution of nearest neighbor distances for our fiducial model compares to the expected distributions with various plausible values of $\ell$ and $M$. We see that the observed distribution is roughly consistent with the distribution we expect for $\ell = 5-6$, and $M=1$, that is, the nearest neighbor distance distribution is about what we would expect if there were typically $\sim 5-6$ library models within the photometric error ellipse of each observation, and if on small scales the distribution of points in photometric space mostly lies along a line. Thus our distribution of nearest neighbor distance is broadly consistent with what we would expect for a well-sampled library drawn from the same distribution as the data.

To summarize, we find that the \cs~libraries are generally in excellent agreement with the observed photometric properties of the LEGUS sample fields. The majority of our sample has $D_5^{\mathrm{norm}} < 1$, meaning that, for the typical observed cluster, there are at least 5 simulated clusters in the \cs~libraries whose magnitudes in all filters are identical to within the photometric errors. We find with no obvious differences in the degree of agreement based on the choice of metallicity, evolutionary tracks, IMF, extinction law, or nebular emission.

\subsection{Posterior PDFs of Star Cluster Mass, Age, and Extinction}

Having verified that our synthetic libraries produce reasonable matches to the data, we now focus on our fiducial library, pad\_020\_kroupa\_MW, and our fiducial prior distribution, $p_{\mathrm{prior}}(\log M, \log T, A_V) \propto M^{-1} T^{-0.5}$, leaving a discussion of the dependence of the results on these choices to the subsequent sections. We analyze the entire photometric catalog described in \autoref{ssec:catalog}, and for each cluster we compute the marginal posterior probability of mass, age, and extinction on a grid of 128 points each, covering the full range of each of these values present in our synthetic library.

The computation is relatively fast -- deriving each posterior PDF requires $\sim 1$ CPU-second per cluster for most high-confidence clusters in the catalog; those with the largest photometric error bars, or that are not well-fit by any models in our catalog (as is the case for many of the visually-rejected candidates, which we have nonetheless analyzed for completeness) may take up to a few tens of CPU-seconds, since large error bars require that we search a larger volume of parameter space. Overall, we find that deriving posterior PDFs for all $\sim 3000$ candidate clusters in the full catalog using a single \cs~library and sets of priors requires a few hours using a multi-core workstation; performing a similar analysis for the $\sim 600$ high-confidence clusters requires tens of minutes.

\begin{figure}
\plotone{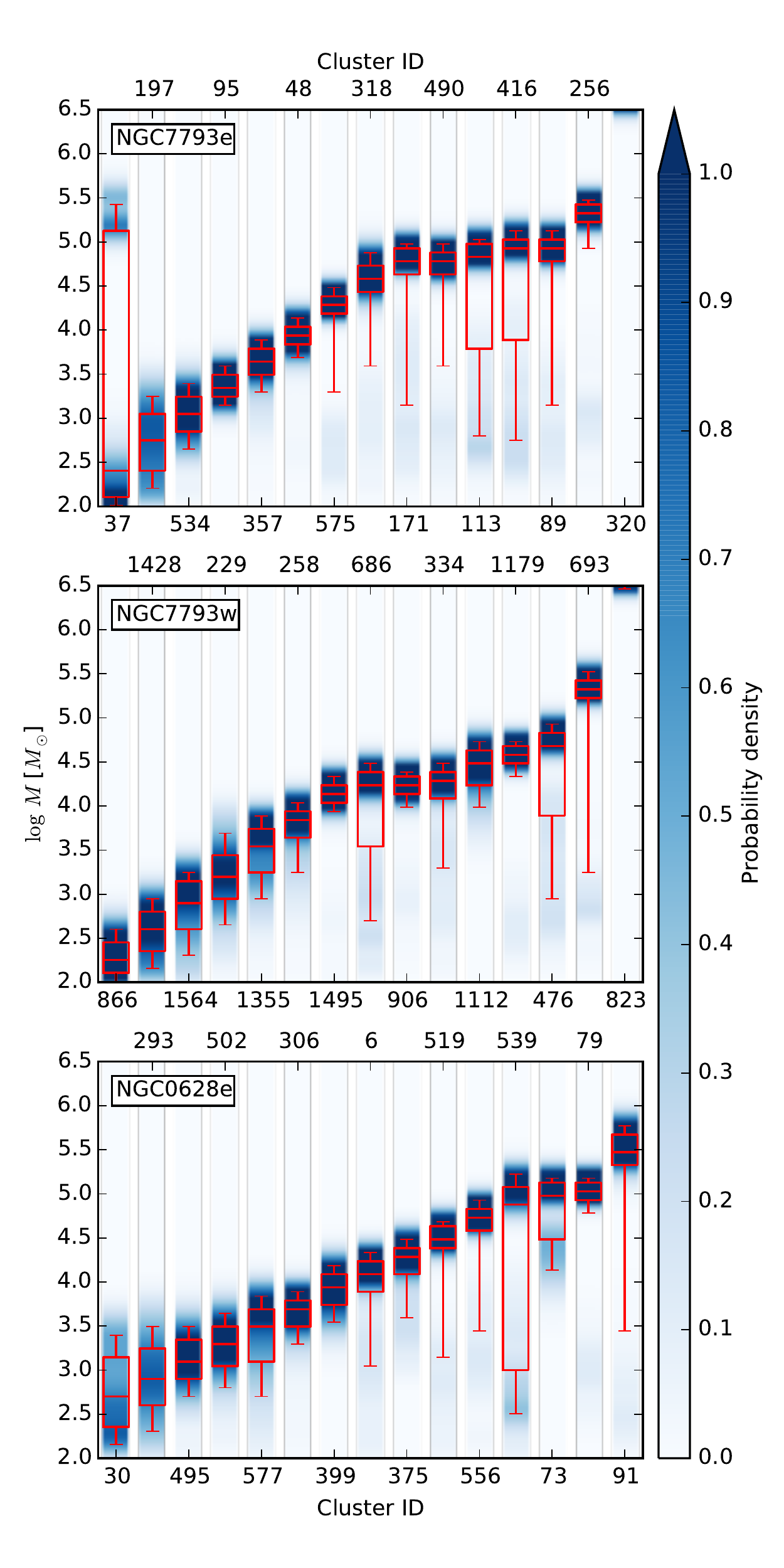}
\caption{
\label{fig:mass_pdfs}
Box and whisker plot showing the marginal posterior probability distributions for star cluster mass for 15 example clusters per field in our 3 sample fields; the cluster IDs in the LEGUS photometric catalog are as indicated, and clusters are ordered from smallest to largest estimated 50th percentile mass. Note that ID numbers appear alternately above and below the boxes and whiskers. For each cluster, the blue colored band shows the probability density at each mass, as indicated in the color bar. Note that the color bar has been clipped above probability densities of 1.0 in order to reveal lower probability density features. The box and whisker plots (red) show percentiles: the lower and upper boxes indicate the range from the 1st to 2nd quartiles, and from the 2nd to 3rd quartiles, respectively. The lower and upper whiskers extend to the 10th and 90th percentiles, respectively.\\
}
\end{figure}

\begin{figure}
\plotone{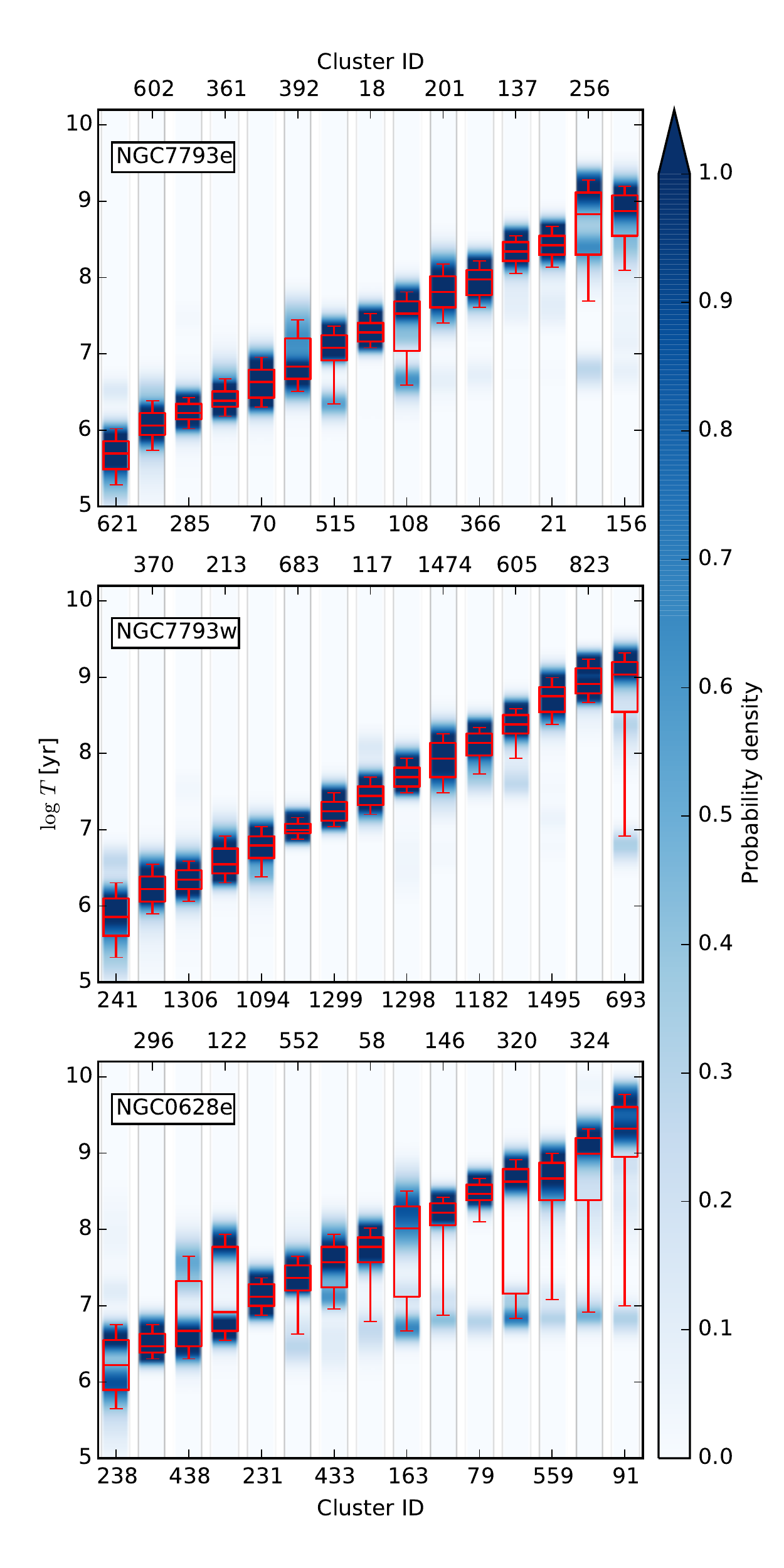}
\caption{
\label{fig:age_pdfs}
Same as \autoref{fig:mass_pdfs}, but now showing the marginal posterior distributions for age. Note that the clusters shown are not the same as the ones shown in \autoref{fig:mass_pdfs}.\\
}
\end{figure}

\begin{figure}
\plotone{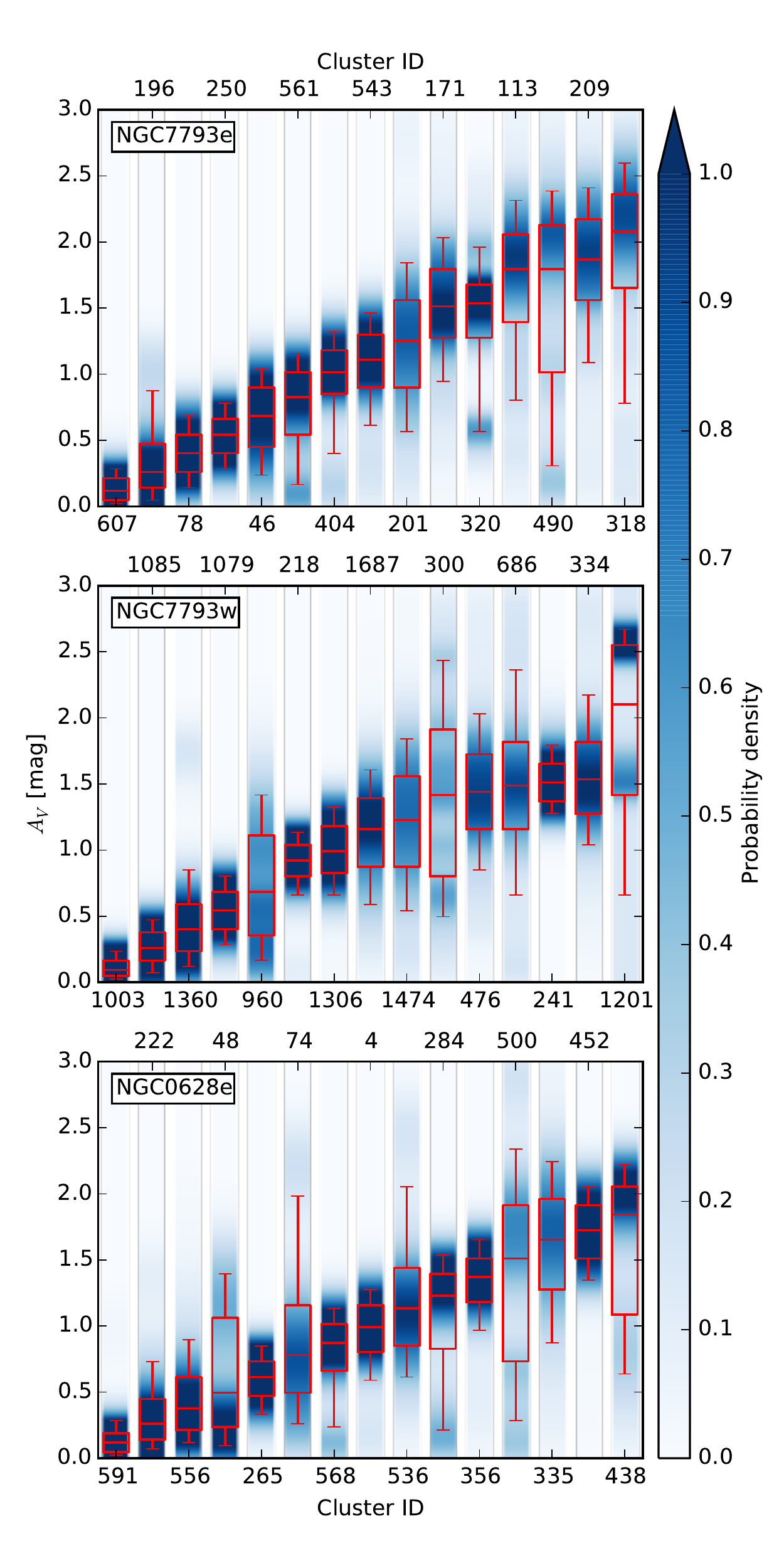}
\caption{
\label{fig:AV_pdfs}
Same as \autoref{fig:mass_pdfs}, but now showing the marginal posterior distributions for visual extinction $A_V$. Note that the clusters shown are not the same as the ones shown in \autoref{fig:mass_pdfs}. \\
}
\end{figure}

Figures \ref{fig:mass_pdfs}, \ref{fig:age_pdfs}, and \ref{fig:AV_pdfs} show sample marginal posterior distributions for mass, age, and extinction for 15 clusters per field in our 3 sample fields. The clusters shown are chosen to be uniformly distributed in the 50th percentile estimates of their log mass, log age, and extinction. From the plots, we can see that in most cases \red{the \cs~models identify a fairly narrow range of possible masses for each cluster}, with a typical interquartile range of $\sim 0.2-0.3$ dex on the posterior probability. The distributions are for the most part unimodal. However, we can see that there are a few cases where the posterior mass distribution is broader or even bimodal, or where there is a tail of probability extending to very different masses, so that the 10th or 90th percentile whiskers extend very far beyond the 1st to 3rd quartile range.

In comparison, the posterior probability distributions of age are somewhat broader and more likely to be bimodal. In the bimodal cases there is often one peak at a relatively old age, and another at an age of $\sim 10^{6.5}$ yr. This is particularly true for NGC\,628, which has the broadest photometric errors. The relative weighting of the two possible age fits, as we shall see below, is not independent of our choice of priors. In these cases the median age may not be a good representation of the actual age, because the median occurs near a local minimum of the PDF that lies partway between the two peaks. The posterior PDFs of extinction are also quite broad. In some cases they are bimodal, while in others they display a single peak but with an extended tail.

\begin{figure}
\epsscale{1.2}
\plotone{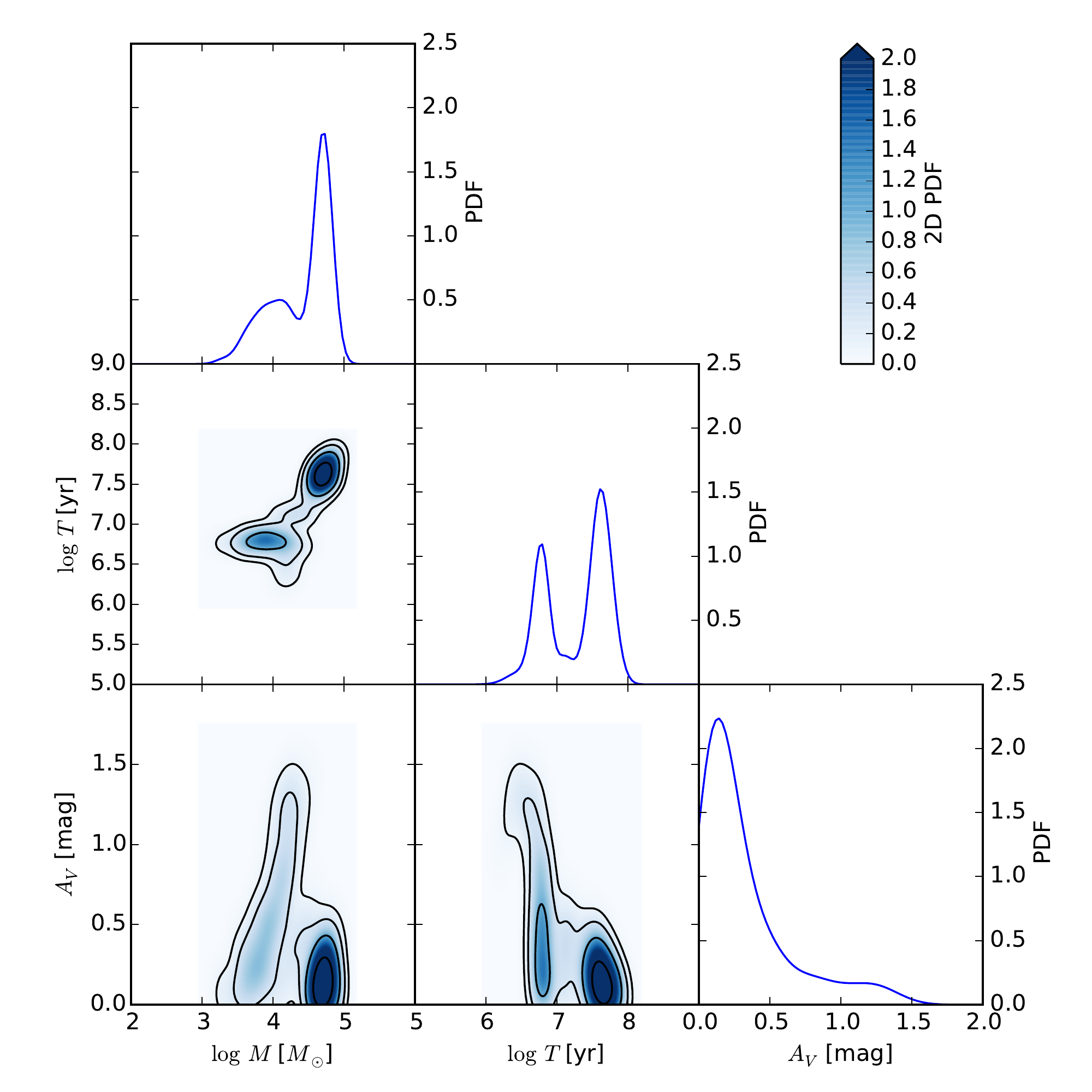}
\caption{
\label{fig:triangle_example}
Triangle plot for an example cluster (ID 56) in NGC\,628e. Line plots show the marginal posterior PDFs for $\log M$, $\log T$, and $A_V$, while raster plus contour plots show the joint marginal posterior PDFs for the joint PDFs of these quantities in combination. Colors indicate probability densities as indicated in the color bar, and contours are spaced in intervals of 0.2. All PDFs are normalized to have unit integral.\\
}
\end{figure}

The nature of these bimodal fits is illustrated in \autoref{fig:triangle_example}, which shows various 2D projections of the posterior PDFs for one example cluster from NGC\,628e, which is typical of many of the bimodal fits we find. As the Figure shows, the data are consistent with two ``islands" of probability. The young island corresponds to an extinction $A_V \sim 0.2 - 1.5$, age $T \lesssim 3-10$ Myr, and mass $M \sim 3000$ $M_\odot$, while the old one is centered near $A_V \sim 0.1$, $T \sim 500$ Myr, $M\sim 3-5\times 10^4$ $M_\odot$. The color and luminosity of this cluster can therefore be fit well by either a relatively massive, extinction-free, old cluster, or a younger, somewhat less massive cluster that is red due to greater extinction.

The conclusions that many clusters when analyzed stochastically show multiple probability maxima is not new, and has been pointed out previously by \citet{fouesneau12a, fouesneau14a}, \citet{de-meulenaer13a, de-meulenaer14a, de-meulenaer15a}, and \citet{krumholz15b}. Indeed, one could obtain such a result even with a deterministic method, provided that one used the full posterior PDF rather than an approximation to it such as a Gaussian centered on the local minimum of $\chi^2$. The primary reason is that there are a number of places in color space where star clusters with disparate physical properties are nonetheless very similar in color. The result is a likelihood function that is not well approximated by a uni-modal Gaussian.

\subsection{Dependence on Choice of Priors}
\label{ssec:priors}

To what extent do the results for the posterior probabilities depend on the choice of prior probability distribution? As noted above, the priors certainly matter for ages young enough that the color provides few constraints, but the priors may also matter in other parts of parameter space as well. To answer this question, we continue to use our fiducial library, but now consider prior probability distributions of the form
\begin{equation}
\label{eq:priordef}
p_{\mathrm{prior}}(\vecx) \propto 
\left\{
\begin{array}{ll}
M^{\beta+1} T, & \log (T/\mathrm{yr}) < 6.5 \\
M^{\beta+1} T^{\gamma+1}, & \log (T/\mathrm{yr}) \geq 6.5 \\
\end{array}
\right.
\end{equation}
with $\beta = -1$ or $-2$, and $\gamma = 0$, $-0.5$, or $-1$; the combination $\beta=-2$, $\gamma=-0.5$ is our fiducial choice.\footnote{Note that the $+1$'s in the exponents in equation (\ref{eq:priordef}) occur because for \cs~we specify the priors on the log of mass and age, while the conventional definitions of $\beta$ and $\gamma$ are in terms of distributions of mass and age, rather than log of mass and age.} That is, we consider cases where we take the prior on mass to be either flat in log mass ($\beta=-1$, all log masses equally likely) or with lower log masses more likely ($\beta=-2$, comparable to what is observed), and where we take the age distribution above $\sim 3$ Myr to be either flat in age ($\gamma=0$), flat in log age ($\gamma = -1$), or intermediate between the two ($\gamma=-0.5$). Flat in age is what would be expected if clusters, once formed, never disrupt, while flat in log age is what would be expected if clusters disperse such that the survival probability is equal for each decade in time. We do not consider variations in the prior on $A_V$, as there seems to be little theoretical or observational motivation to do so. The effects of varying the prior on $A_V$ are likely degenerate with the effects of varying the prior on age.

\begin{figure*}
\epsscale{1.2}
\plotone{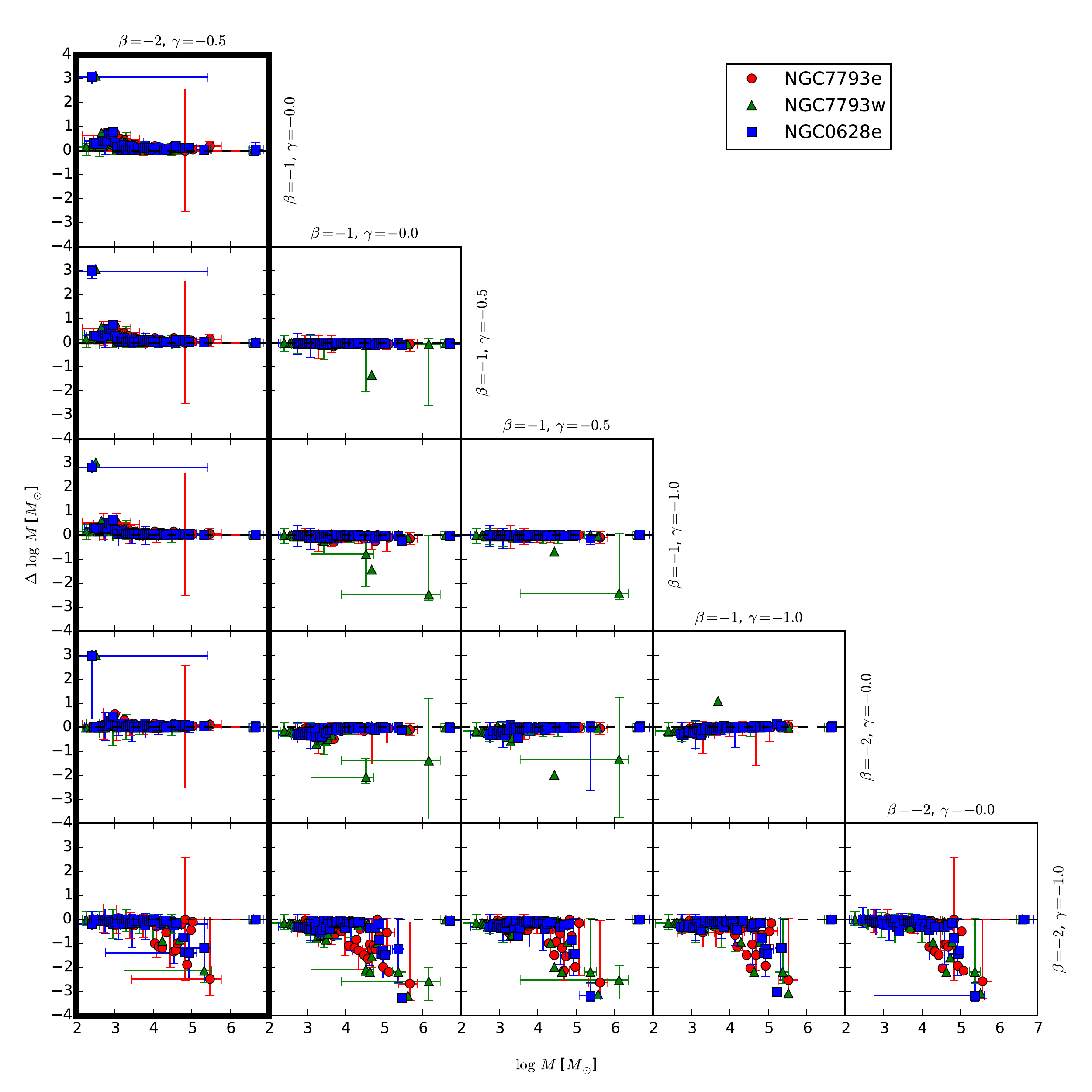}
\caption{
\label{fig:prior_mass}
Comparison of how the posterior masses we derive for the star clusters depend on our choice of priors. In each panel, the $x$ axis shows the log of mass $(\log M)_x$ derived with one choice of prior (characterized by $(\beta, \gamma)$, the slopes of the mass and age distributions -- see \autoref{eq:priordef}), while the $y$ axis shows the difference $\Delta \log M = (\log M)_y - (\log M)_x$ between this value and the 50th percentile derived using a different prior. In each column, all panels use the same prior to generate the value $(\log M)_x$ shown on the $x$ axis, as indicated by the label at the top of the column; in each row, all panels use the same prior to generate $(\log M)_y$ and thus $\Delta \log M$, as indicated by the labels at the right of the rows. The leftmost column, highlighted, uses our fiducial choice $(\beta=-2, \gamma=0)$ for $(\log M)_x$. Symbols of different shapes and colors correspond to different galaxies, as indicated in the legend, and the dashed horizontal line shows $\Delta \log M = 0$, indicating that the results are independent of the prior. Finally, the point plotted for each cluster is the inferred 50th percentile mass, while the horizontal and vertical error bars show the range from the 10th to 90th percentile as inferred for each set of priors. For the vertical error bars, the range in $(\Delta \log M)$ plotted is the range in $(\log M)_y$ only, rather than reflecting a composite of $(\log M)_x$ and $(\log M)_y$. We show the 10th to 90th percentile range for only a subset of the data in order to minimize clutter.
}
\end{figure*}

\begin{figure*}
\epsscale{1.2}
\plotone{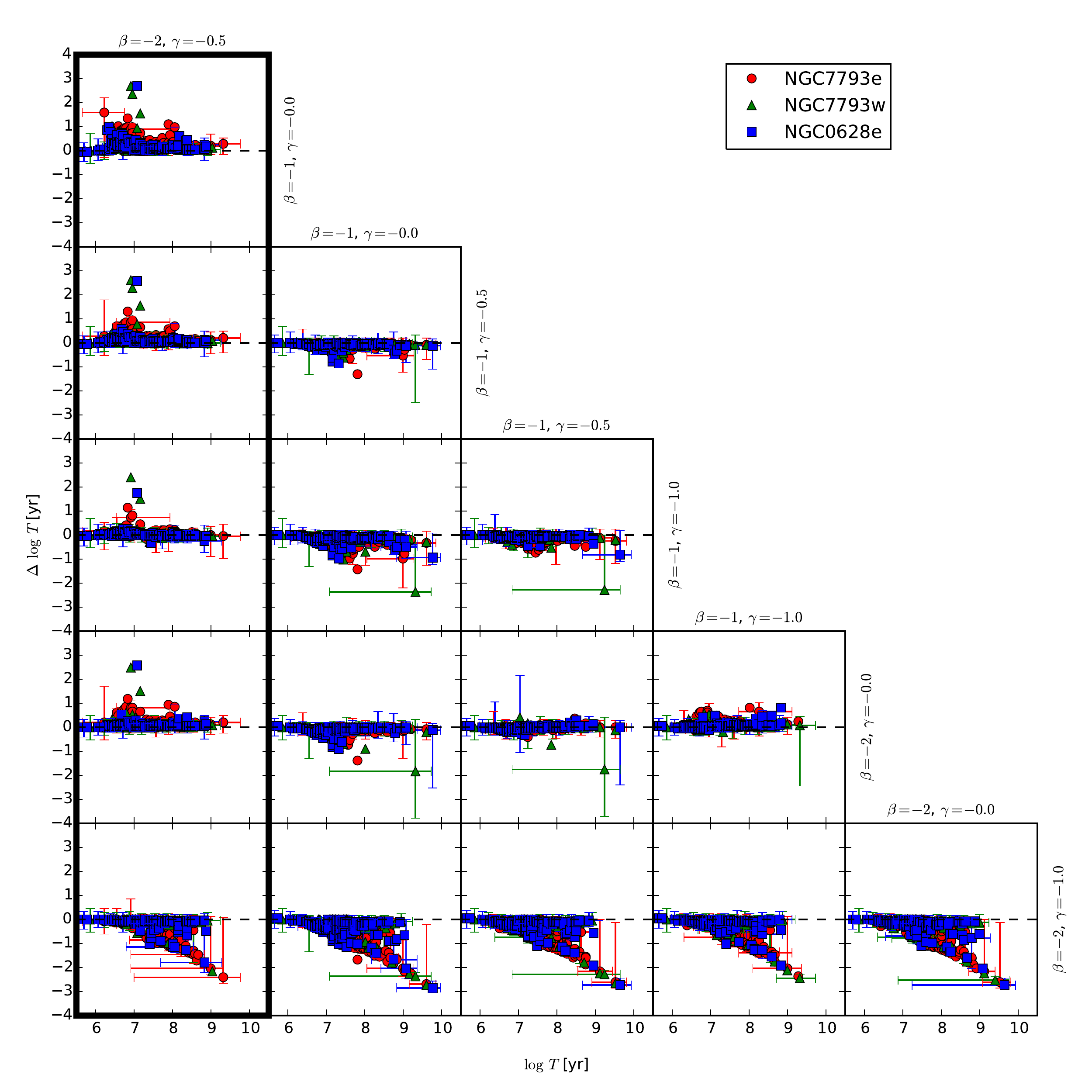}
\caption{
\label{fig:prior_age}
Same as \autoref{fig:prior_mass}, but showing a comparison of ages rather than masses derived using different priors.
}
\end{figure*}

In \autoref{fig:prior_mass} and \autoref{fig:prior_age}, we show how the masses and ages we derive for our star clusters depend on our choice of prior. We see that the choice of prior has relatively little impact on the results for most cluster masses, typically moving the median by an amount significantly less than the 10th to 90th percentile range. The prior that deviates most from the others is $\beta=-2$, $\gamma=-1$. With this choice the majority of clusters are still largely unchanged, but a substantial tail of clusters appears for which the deviation $\Delta\,\log M$ is substantially less than zero, indicating that the prior $\beta=-2$, $\gamma=-1$ leads us to a substantially smaller mass estimate than alternative choices.
The effect of prior choice on the inferred ages is somewhat greater, but as with masses, the only case that shows a very significant variation is $\beta=-2$, $\gamma=-1$, which again produces a tail of clusters with $\Delta \log T$ substantially negative. In most cases this is simply an understandable shift in power between two peaks of a bimodal PDF. As in the example shown above in \autoref{fig:triangle_example}, in some cases there are two islands of probability that both fit the observed photometry reasonably well. In these cases a change in priors will enhance one island and suppress the other. Not surprisingly, in these cases the median shifts considerably.

However, there are also a number of extreme outliers, where the amount by which the posterior PDF shifts when we change our priors is so great that the previously inferred median now lies outside the 10th to 90th percentile range. Many of these points represent clusters with substantial photometric errors, clusters that are not well-fit by our model libraries (perhaps because they are composites of multiple populations of different ages), or both. It is not surprising that the derived results for these clusters should be very sensitive to the choice of prior. When the error bars on the observations are large, the likelihood ratio is close to flat, and thus the posterior probability distribution is little altered from the prior one. Something analogous occurs if no model cluster in our library is a good fit to the observations, and instead there are a broad range of models that are less accurate fits.

\begin{figure}
\epsscale{1.2}
\plotone{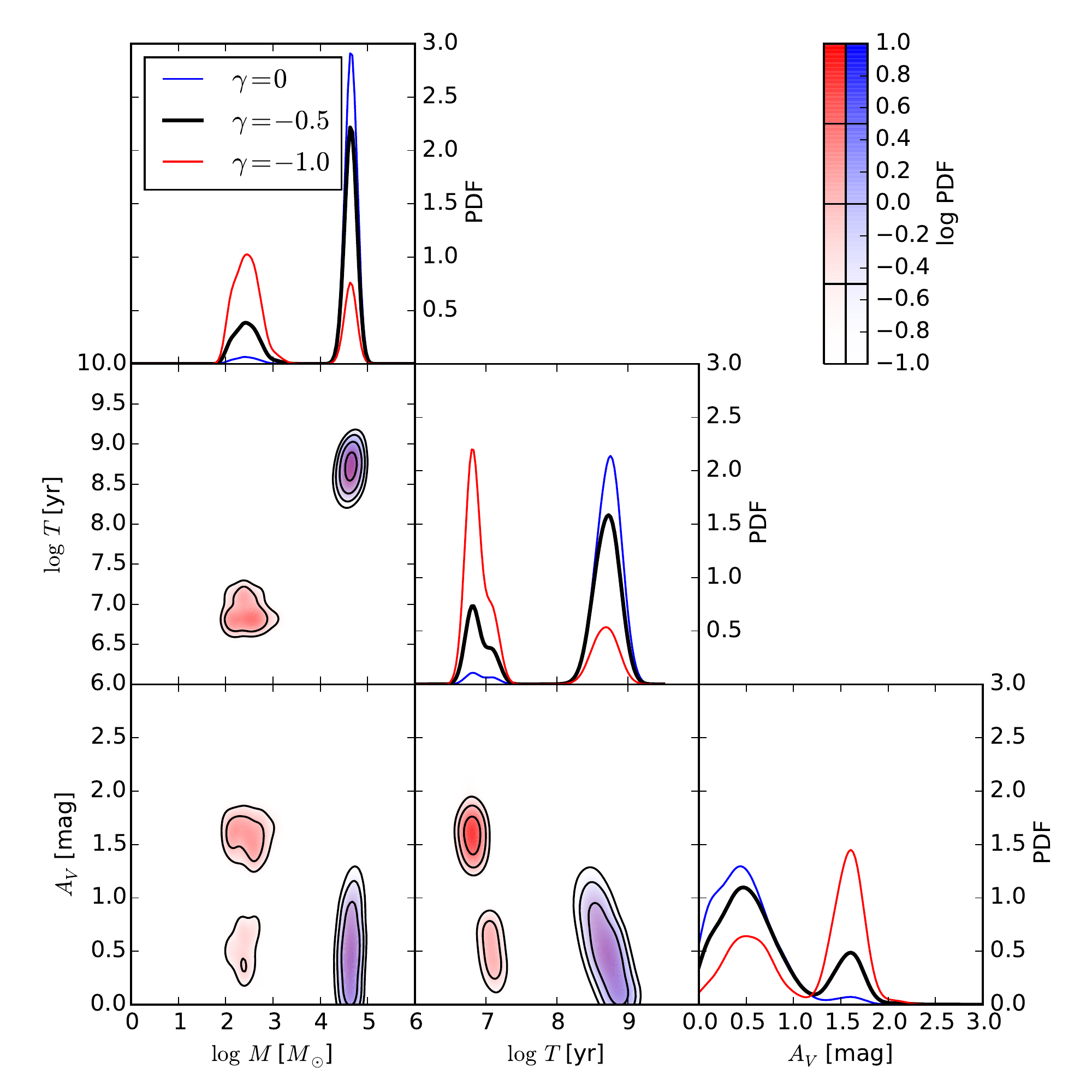}
\epsscale{1.0}
\caption{
\label{fig:triangle2}
Triangle plot for an example cluster (ID 320) in NGC\,628e. Panels are similar to those in \autoref{fig:triangle_example}. In the panels along the central diagonal showing the 1D marginal PDFs, the black line corresponds to the fiducial case ($\beta=-2$, $\gamma=-0.5$), while the blue and red lines refer to the alternate priors $\beta = -2$, $\gamma=0$ and $\beta = -2$, $\gamma=-1$, respectively. In the three lower left panels, blue and red colors show the 2D PDF on a logarithmic intensity map (as indicated in the colorbar) for $\beta = -2$, $\gamma=0$ and $\beta = -2$, $\gamma=-1$, respectively. Black contours show the 2D PDF for the fiducial case, with contours placed at log probability density values starting at $-1$ and increasing by 0.5 per contour. The 1D and 2D PDFs are all normalized to have unit integral.\\
}
\end{figure}

However, there is also a population of clusters without large photometric error bars, and that are well-fit by our model library, that nevertheless show very substantial changes in their posterior PDFs depending on our choice of prior. To understand what is happening in these cases, in \autoref{fig:triangle2} we show the 2D posterior PDF for an example cluster whose posterior PDF changes substantially depending on the choice of prior, but that does not have unusually large photometric errors, and that is well-fit by our models. As the plot shows, this cluster also has two islands of probability, one centered at a mass of $\sim 10^{4.5}$ $M_\odot$ and an age of $\sim 500$ Myr, and a second centered at a narrow range of ages $\sim 5-10$ Myr and masses of $10^{2} - 10^{3}$ $M_\odot$. Unlike the degeneracy between age and extinction shown in \autoref{fig:triangle_example}, these two possibilities need not sit at substantially different $A_V$ values. Instead, the degeneracy take a rather different form, which is more closely related to IMF sampling.

The older age possibility corresponds to typical colors and luminosities for clusters of that mass and age range. On the other hand, the young age case corresponds to clusters that are on the far tail of the luminosity and color distributions for that age and mass. Thus the young age case represents extremely unusual photometric properties for clusters so young and low mass to have, resulting from very improbable draws from the IMF. If all ages are considered equally likely, then the rare young, low-mass cases are rejected as unlikely. However, the difference between $\gamma=-1$ and $\gamma=0$, as indicated by red and blue colors in \autoref{fig:triangle2}, amounts to the difference between a prior that says that there should be $\sim 100$ times as many clusters with ages from $5-10$ Myr as clusters with ages of $505-510$ Myr, and a prior that says there should be roughly equal numbers of clusters in those two ranges. If our prior is the former rather than the latter, then our Bayesian estimate assigns comparable total probabilities to the two possible fits, leading to a very different posterior PDF.

It is an open question to what extent the dependence on the choice of prior could be reduced or removed by the availability of H$\alpha$ data, which is particularly sensitive to the youngest ages and therefore good at discriminating between otherwise-degenerate models. \citet{fouesneau12a}, analyzing star clusters in M83, found that H$\alpha$ (as measured in their case by \textit{HST}'s F657N filter) was very helpful in breaking degeneracies between fits. However, their analysis did not make use of F275W and instead had F336W as its bluest filter. The extra UV coverage provided by LEGUS's F275W band should provide at least some of the same sensitivity to very young ages that \citeauthor{fouesneau12a}~obtained from their H$\alpha$ data. Moreover, \citeauthor{fouesneau12a}'s sample was limited to clusters with masses above $\sim 10^3$ $M_\odot$, where stochastic effects are somewhat less important than in our sample. Since H$\alpha$ is produced primarily by the most massive stars, it is particularly vulnerable to stochastic effects, which might reduce its ability to discriminate between models for our data set. In any case, a survey to obtain H$\alpha$ data for a subset of LEGUS galaxies is underway (HST-GO-13773, PI: R.~Chandar). As those data become available we will re-run our analysis pipeline using them, which should provide an answer to this question.

It may also be possible to break degeneracies and remove dependence on the choice of prior using other discriminators. For example, the amount of scatter in surface brightness within the aperture \citep{whitmore11a} and the number of red stars in the vicinity of the cluster \citep{kim12a} have both been suggested as age-indicators. At present it is not clear how to build these into a Bayesian framework such as the one we have developed, as this would require extension of the likelihood function to include this information.

\subsection{Dependence on Choice of Tracks, IMF, Metallicity, Extinction Curve, and Nebular Emission}

\begin{figure}
\epsscale{1.2}
\plotone{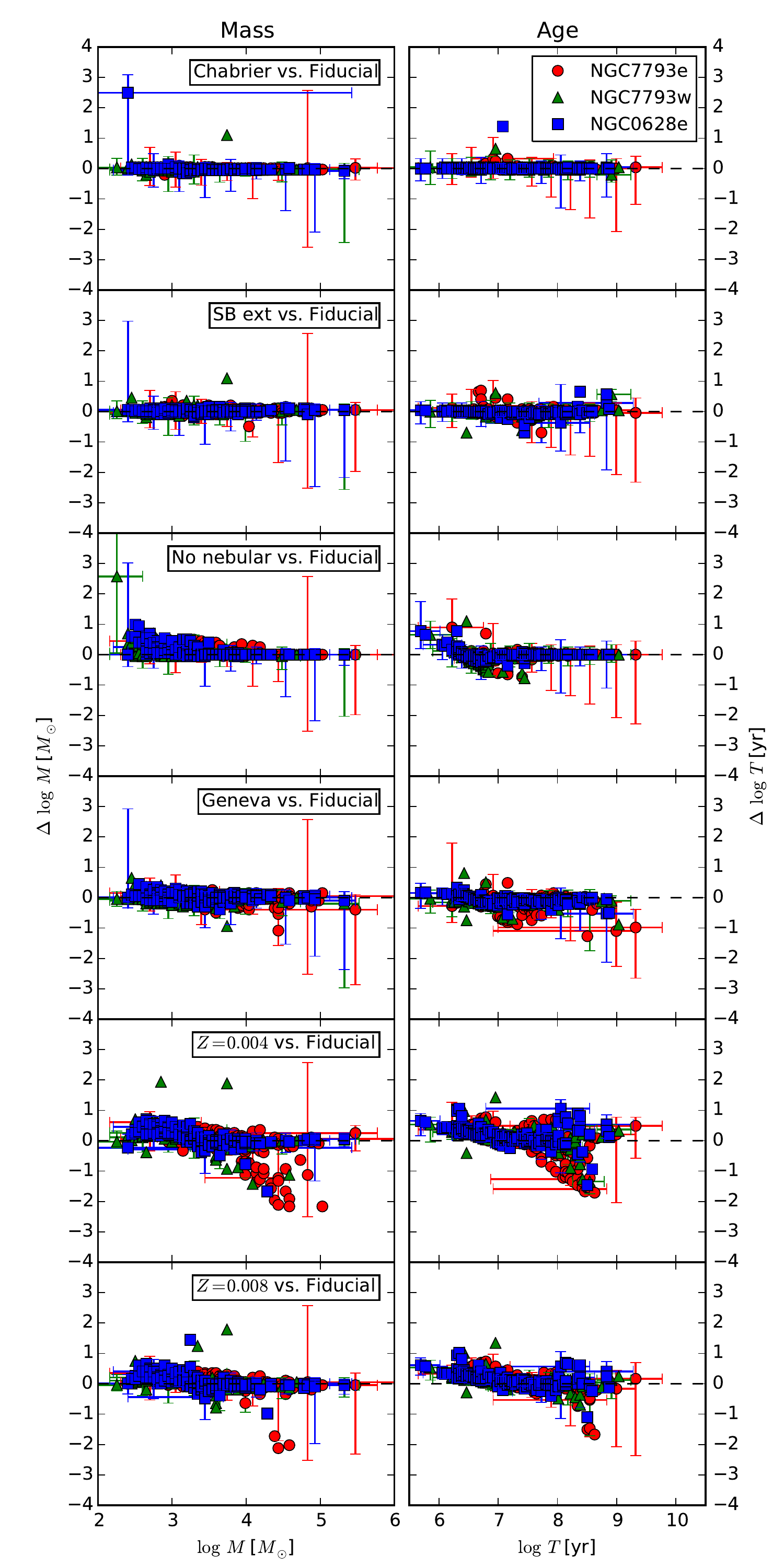}
\epsscale{1.0}
\caption{
\label{fig:modelcomp}
Comparison of posterior PDFs derived using our fiducial model, pad\_020\_kroupa\_MW, and various alternatives: from top to bottom, pad\_020\_chabrier\_MW (\citet{chabrier05a} IMF instead of \citet{kroupa01a} IMF), pad\_020\_kroupa\_SB (starburst instead of Milky Way extinction curve), pad\_020\_kroupa\_MW\_noneb (same as the default library, but omitting nebular emission), gen\_014\_kroupa\_MW (Geneva stellar tracks at $Z=0.014$ versus Padova tracks at $Z=0.02$), pad\_004\_kroupa\_MW ($Z=0.004$ instead of $Z=0.02$), and pad\_008\_kroupa\_MW ($Z=0.008$ instead of $Z=0.02$). In each panel, the fiducial model is on the $x$ axis, and the comparison is on the $y$ axis. The left column shows masses, and the right shows ages. Points plotted are the 50th percentile estimates of each quantity, and error bars (plotted for only a subset of the points to reduce clutter) indicate the 10th to 90th percentile range. Different colors and plot symbols correspond to different fields, as indicated in the legend.
}
\end{figure}

We next examine the extent to which our results depend on our choice of tracks, IMF, metallicity, and extinction curve. In \autoref{fig:modelcomp} we show a comparison between results derived using our fiducial model, pad\_020\_kroupa\_MW, and results derived using models with different extinction curves, metallicities, IMFs, and stellar tracks, and omitting nebular emission. In all these comparisons we use our fiducial priors, $\beta=-2$, $\gamma=-0.5$, but the results are comparable for other priors provided that we use the same prior for each library. Examining the figure, it is clear that the choice of IMF and extinction curve make almost no difference to the final posterior PDF. This is to be expected. For the IMF, the \citet{chabrier05a} and \citet{kroupa01a} IMFs we have tried differ mostly at the brown dwarf end, which has little impact on the integrated light properties even for old stellar populations. They also differ in that one is truncated at 100 $M_\odot$ and the other at $120$ $M_\odot$, but those stars appear to be rare enough that the difference made to the integrated light by including or omitting them is smaller than the level of variation induced by IMF sampling effects. As for $A_V$, recall that the $A_V$ values we infer are generally modest. If there is little extinction, then the shape of the extinction curve matters little. The choice of IMF also makes little difference, which is, again, not surprising. 

Whether we include nebular emission or not only makes a difference at young ages and low masses, though the latter is a selection effect -- our flux limit is such that our low-mass clusters are exclusively young. In this range, we find that models excluding nebular emission produce systematically higher masses. This result is easy to understand: if we ignore the light produced by the nebula, then a higher stellar mass is required to produce the observed light. The effect of ignoring nebular emission on ages is more subtle, and again points to the importance of priors. We find that models excluding the nebular light do not produce 50th percentile ages below $\approx 3$ Myr, while our fiducial models show no such exclusion. The result for the no-nebular case is driven by the fact that the colors of stars are essentially constant at ages $\lesssim 3$ Myr, so the likelihood function we compute by comparing to the observed colors is flat over this age range. Combined our prior that all ages $T$ below $10^{6.5}$ yr are equally likely, the posterior distribution with respect to $\log T$ must therefore peak at larger $T$. On the other hand, when including nebular emission colors do evolve, at least mildly, at younger ages, and thus the likelihood function is not flat and we differentiate ages below $3$ Myr. As a result, the 50th percentile values populate the full range of ages $\lesssim 3$ Myr.

The choice of tracks has modest but non-negligible effects. Compared to our fiducial case, the Geneva tracks seem to favor somewhat lower masses and ages. However, it is important to notice that the changes are largest for those clusters that have very significant uncertainties, so that the one-to-one line still falls mostly within the 10th to 90th percentile range. In effect, in those cases where the colors are ambiguous and the posterior PDF is multi-peaked, changing from one set of tracks to another can favor or disfavor one of the two peaks.

The largest effect is from metallicity. Compared to Solar metallicity, the lower metallicity models favor substantially lower ages and masses for intermediate mass and age clusters, and slightly higher masses and ages for the youngest and lowest mass clusters. The effect at young ages and low masses appears to be similar to that seen in the the no-nebular versus nebular comparison. This results from a complex series of effects of metallicity on the color: lower metallicity increases the ionizing flux and raises the nebular temperature, thereby increasing continuum and hydrogen recombination line emission. On average it also weakens emission from metal lines, though some lines that are particularly temperature-sensitive (e.g.,~[O~\textsc{iii}] $\lambda5007$) may also get brighter. The resulting shift in colors appears to favor older ages and higher masses for the youngest and lowest mass clusters. The effect at higher mass and age comes largely from the interaction of colors with IMF sampling effects. Choosing a low metallicity tends to drive the model colors to the blue, beyond the point where they agree well with the observations. To match the observed colors requires driving the colors back to the red, and one way of doing this is to have a lower mass cluster that under-samples the massive end of the IMF, thereby producing redder colors.

Beyond the individual model comparisons, perhaps the most striking result shown in \autoref{fig:modelcomp} is how little difference the choice of models makes. The only one of the variations that we have examined that appears to make a substantial difference is using $Z=0.004$, and, at the youngest ages perhaps, including or not including nebular emission. The primary reason for this insensitivity is that widths of the posterior PDFs for age and mass estimates are sufficiently large that they swamp systematics such as the choice of extinction curve, IMF, evolutionary tracks, and, at least over a limited range, metallicity. Perhaps these choices would become important if we had additional observational constraints beyond the five photometric bands to which we currently have access, but at present they are not the limiting factors in our knowledge.

\subsection{Comparison of Stochastic and Non-Stochastic Models}
\label{ssec:ygcomp}

Our final analysis step is a comparison of the results we obtain using \cs~stochastic models to those we get from the deterministic \yg~code. For this comparison, we use our fiducial library, pad\_020\_kroupa\_MW, and prior distribution, $\beta=-2$, $\gamma=-0.5$. We can investigate this question on two levels. We can first ask how well \cs~and \yg~models match the observed photometry, and whether one provides better matches of the observations than the other. We can then ask how the results we derive for cluster properties, and the nominal errors on those results, compare for the two methods.

\subsubsection{How Well Do \cs~and \yg~Reproduce the Observed Photometry?}

\begin{figure}
\epsscale{1.2}
\plotone{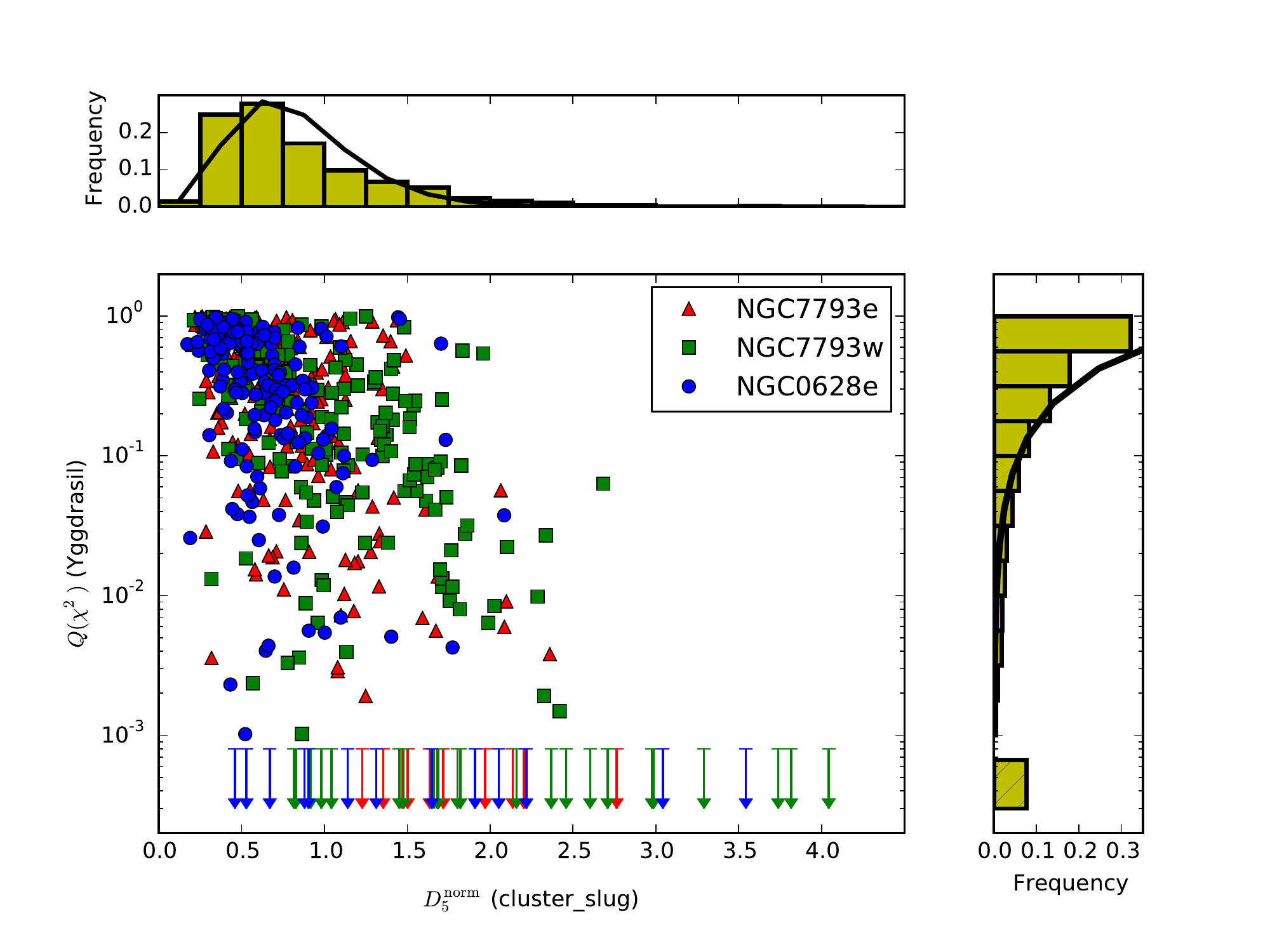}
\epsscale{1.0}
\caption{
\label{fig:cs_yg_errordist}
Distribution of errors between the observed photometry and the models. In the central panel, for each cluster the $x$ coordinate shows $D_5^{\mathrm{norm}}$ (\autoref{eq:photdistnorm}), while the $y$ coordinate shows $Q(\chi^2)$. Values of $Q < 10^{-3}$ are shown as upper limits, indicated by arrows. The histograms in the flanking panels show the binned distribution of $D_5^{\mathrm{norm}}$ (top panel) and $Q$ (right panel) values for all clusters; the lowest, hatched bar in the frequency distribution for $Q$ shows the frequency of points with $Q < 10^{-3}$. In both histograms, the solid lines indicate the distribution of values we would expect for \cs~or \yg~models that provide good fits to the data -- see main text for details.\\
}
\end{figure}

To investigate the consistency or lack thereof between the \cs~and \yg~models and the observed photometry, we must first develop a statistic to quantify it. For \cs, we parameterize the quality of agreement between the observations and the library using $D_5^{\mathrm{norm}}$, the normalized photometric distance between an observed cluster and its 5th closest neighbor in the \cs~library (\autoref{eq:photdistnorm}), exactly as in \autoref{ssec:consistency}. For $\yg$, the best fitting model is determined via a $\chi^2$ minimization procedure. From the value of $\chi^2$ determined for the best fit (and the number of degrees of freedom in the model), we can compute the $Q$ statistic $Q(\chi^2)$, which formally is the probability that the value of $\chi^2$ would exceed the measured one even if the model were correct. Thus for example $Q(\chi^2) = 0.3$ means that, even if the best fit model were true, there is a 30\% chance that each observation of that cluster would yield photometry for which the $\chi^2$ value would exceed our measured value. Thus values of $Q(\chi^2)$ near unity correspond to good fits, while ones with $Q(\chi^2) \ll 1$ indicate either that the error bars have been underestimated or that the best fitting model in the \yg~library is probably not a good fit to the data.

\autoref{fig:cs_yg_errordist} shows the distribution of $Q(\chi^2)$ versus $D_5^{\mathrm{norm}}$ for our photometric catalog. In this plot, good fits for \cs~lie on the left side of the diagram, while poor fits lie on the right. Similarly, good fits for \yg~lie at the top of the plot, and poor fits toward the bottom. The plot shows, consistent with \autoref{fig:error_dist}, that data are generally well-reproduced by the \cs~library. The solid line in the histogram of $D_5^{\mathrm{norm}}$ models shows the expected distribution for a library with $\ell = 6$ models within an observational error circle and $M=1$ dimensional data (see \autoref{ssec:consistency}); it is simply the binned, differential version of the cumulative distribution shown in \autoref{fig:error_dist}. The \cs~distribution of errors is roughly consistent with this distribution, confirming that most clusters have good \cs~fits.

The same is not true of the \yg~models. For a good fit the $Q(\chi^2)$ statistic should be distributed so that a fraction $f$ of models have $Q(\chi^2) < f$. In fact, \autoref{fig:cs_yg_errordist} shows that there is a significant excess models at $Q \ll 1$. The solid line in the histogram of $Q$ values shows the distribution we would expect for a good fit, and there are clearly more small values of $Q$ than expected, and significantly fewer values close to unity. In particular, nearly 10\% of models have $Q < 10^{-3}$, while for a good fit such large discrepancy should occur $0.1\%$ of the time. In contrast, $<5\%$ of clusters have $D_5^\mathrm{norm}>2$, and only 1\% have $D_5^\mathrm{norm}>3$.

One possible explanation for the excess of clusters with $Q \ll 1$ is that our observed catalog includes some objects that are not truly simple stellar populations, for example because they are blends of two clusters of different ages. However, we can immediately reject this as the dominant explanation, because such clusters should also be fit poorly by \cs~models, placing them in the lower right corner of the main panel of \autoref{fig:cs_yg_errordist}. While there are a few objects of this sort, there are also a very large number of clusters for which \yg~returns a $Q$ value of $\sim 10^{-2}$ or less, but \cs~nevertheless finds at least 5 simulated clusters whose photometry matches the observed values within 1 or 2 times the photometric errors. Indeed, even for the subset of clusters for which \yg~produces $Q<10^{-2}$, the mean value of $D_5^{\mathrm{norm}}$ is $1.8$. This is despite the fact that \yg~is only fitting the colors (since the mass and thus the absolute luminosity are left as free parameters), while \cs~is fitting both the colors and the absolute magnitude. In contrast, there is no corresponding population of clusters in the upper right part of the diagram, which would indicate that \yg~finds a good fit but \cs~does not.

Many of the clusters for which \yg~does not find good fits are those for which \cs~assigns relatively low masses; the mean of \cs's 50th percentile mass estimate for the subset of clusters for which \yg~finds $Q<10^{-2}$ is 1000 $M_\odot$, smaller than the 50th percentile mass for the entire sample by a factor of several. This strongly suggests that much of the problem is driven by \yg's assumption of a well-sampled IMF, which fails in the low mass clusters. For these cases, exploring the full range of stochastic color variation is crucial to finding a good match to the data. However, there also some relatively massive clusters for which \cs~finds a good match but \yg~does not. These may be cases where, despite the cluster's overall relatively large mass, its colors are still significantly influenced by a small number of stars undergoing short-lived phases of stellar evolution. For example, a small number of  He core stars undergoing blue loops might dominate the UV light budget of an otherwise old and red stellar population. For these rare phases of stellar evolution, the continuous assumption used in \yg~may fail even for relatively massive clusters.

\subsubsection{Comparison of \cs~and \yg~Best Fits and Errors}

\begin{figure}
\plotone{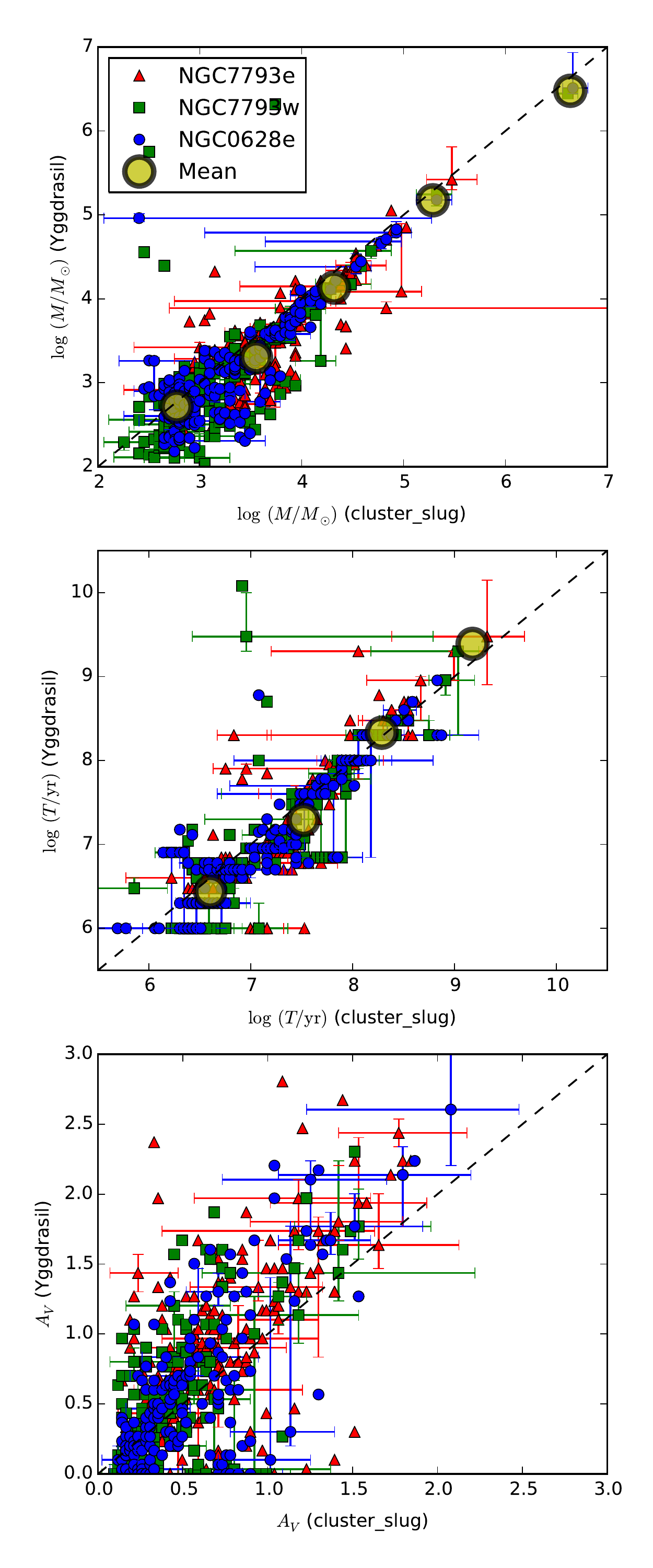}
\caption{
\label{fig:ygcomp}
Comparison of cluster masses (top panel), ages (middle panel), and extinctions (bottom panel) computed with \yg~and \cs. In all panels, the central point is plotted at the 50th percentile value returned by \cs~on the $x$ axis, and the best fit returned by \yg~on the $y$ axis. Error bars, which we show only on a subset of points for clarity, indicate the 68\% confidence interval for \yg, and the 16th to 84th percentile range for \cs. Point colors and shapes indicate the field for each cluster, and black dashed lines show the $1-1$ relation. In the top two panels, large yellow points show the mean logarithmic mass and age, respectively, for both $\cs$ and $\yg$. Mean masses are computed by averaging the clusters in bins of 50th percentile \cs~mass from $\log (M/M_\odot) = 2-3$, $3-4$, $4-5$, $5-6$, and $6-7$; ages are computed via an analogous procedure over bins from $\log (T/\mathrm{yr}) = 6-7$, $7-8$, $8-9$, and $9-10$. In the middle panel, the alignments of the \yg~fits at a set of common ages is an artifact of the fitting procedure.
}
\end{figure}

Having discussed how well the models libraries match the data, we now ask how well the predicted cluster properties and the errors on those properties agree between the two methods. \autoref{fig:ygcomp} shows a comparison of the best-fit \yg~values to the 50th percentile \cs~values. This is not quite a fair comparison, since in some cases the 50th percentile \cs~mass does not actually lie near a probability maximum. Nonetheless, the obvious alternative, plotting the \cs~point at the probability maximum, is no better. Some imprecision is inevitable when comparing a full posterior PDF to a single best fit.

Comparing masses in \autoref{fig:ygcomp}, we notice that the models generally agree fairly well at high masses, but at low masses the \yg~models on average tend to produce lower masses compared to \cs's 50th percentile, making the scatter about the $1-1$ line somewhat asymmetry. Despite this, the mean masses for the population as a whole (indicated by the yellow circles in \autoref{fig:ygcomp}) are quite similar. The origin of this effect is almost certainly IMF under-sampling. For $\sim 500$ $M_\odot$ clusters, the most common outcome by number is that the cluster will under-sample the massive end of the IMF, and thus will be under-luminous compared to what would be expected for a fully-sampled IMF. Because \yg~assumes full sampling it assumes a mass to light ratio that is too low, and ends up with a mass estimate that is too low as well. In contrast, \cs~correctly accounts for the sampling effect and assigns a broad PDF whose median value lies at higher masses than \yg's best fit. The effect begins to appear at masses below roughly $10^{3.5}$ $M_\odot$, consistent with earlier analyses \citep[e.g.,][]{elmegreen00b, cervino04a, cervino06a, da-silva12a, fouesneau14a}. However, when averaging over a large population of clusters, there will be a few that happen to be over-luminous rather than under-luminous for their mass, so that the mean is the same as for fully-sampled IMF. Consequently, while there is an asymmetric bias in the ages of individual clusters, estimates for the mean of the entire population as much less affected.

There are also systematic offsets in \yg~and \cs's age and $A_V$ distributions. Here the phenomenology is more complex. As we have seen, the ages one derives from \cs~are not independent of the choice of priors. There are often genuine ambiguities, with multiple age-$A_V$ combinations providing plausible fits to the data. The 50th percentile value that \cs~derives will depend on how the priors weight these two reasonably good fits. The best-fit procedure used by \yg~should be roughly equivalent to searching for the highest peak in age-$A_V$ space, based on priors that are flat in $\log T$ and $A_V$. The default \cs~priors used for the results shown in \autoref{fig:ygcomp} have priors that are flat in age rather than log age (though this is partly offset by the prior to low masses, which implicitly favors younger ages), and so it is not surprising that they produce somewhat different results. Despite the offsets, however, we see that the population means are again fairly similar between \cs~and \yg.

The greatest discrepancy would appear to be in the values of $A_V$ derived by the two methods. However, this is somewhat misleading, as the error bars on $A_V$ are often extremely large. Thus despite the seeming large offset between the \cs~medians and \yg~best fits, in reality almost all of the data points have the $1-1$ lines within their allowed error range.

Finally, we note that the uncertainty range produced by \cs~is in almost all cases larger than the one produced by \yg. This is particularly true of masses at the low mass end, where \cs's 68\% confidence range is often close to an order of magnitude wide, while \yg's is so small that it is nearly invisible in \autoref{fig:ygcomp}. This is likely because \yg's error estimate only includes the formal error coming from propagation of the photometric errors though the deterministic model grid, while \cs~properly captures the additional uncertainties coming from stochasticity and from degeneracies. Clearly the latter two effects dominate the total error budget, leading \cs~to have much larger output errors than \yg.

\section{Properties of the Cluster Populations}
\label{sec:popresults}

Having analyzed the properties of the individual clusters and their statistical and systematic uncertainties, we now seek to extent our analysis to the properties of the cluster population as a whole. Before commencing this exercise, we note that the results we derive in this section should \textit{not} be taken as representative of the underlying properties of star clusters in our sample galaxies. Instead, we seek to derive the properties of those clusters that have made it into the LEGUS photometric catalog, which is the convolution of the true distribution of properties with the catalog selection. To convert these to intrinsic properties our analysis would have to be corrected for completeness, and such a correction is beyond the scope of the present work. In general we note that, because our catalog consists of visually-inspected clusters, and the main criterion for being subject to visual inspection is exceeding a magnitude limit, our sample should be closer to magnitude-limited than mass-limited. This will tend to result in a steeper age distribution than is present in reality \citep[e.g.,][]{lamers09a}.

\subsection{Methods}

Our goal here is to derive the probability distribution functions for ages and masses of clusters, or their joint distribution, given the posterior PDFs for each individual cluster that we have derived in the previous section. Deriving this requires some care. The traditional method to derive the properties of the population, when each cluster has been fit using a traditional $\chi^2$ approach, is to create bins in mass and age, and then assign clusters to the bin corresponding to the best fit. The errors on each bin can then be computed from Poisson statistics.\footnote{A slightly more accurate procedure is to distribute the clusters between bins based on the convolution of the bin with their formal uncertainties, but since we have already seen that traditional $\chi^2$ method greatly underestimates the uncertainties (\autoref{ssec:ygcomp}), this is only a marginal improvement.} This is the method we will use to analyze the \yg~results below. As usual, one must make a choice of how to place the bins, and we will consider two cases: one where the bins are placed uniformly in the log of mass or age, and one where they are placed so as to have a fixed number of clusters per bin. Note that the latter choice also maintains a fixed value of the relative error on the PDF: since the error for a bin containing $N$ clusters is $\sqrt{N}$, the ratio of the error to mean is just $1/\sqrt{N}$, and thus a fixed value of $N$ implies a fixed fractional error.

This method is clearly not well-suited to our posterior PDFs, which are broad and multiply-peaked. Moreover, we wish to avoid binning, since binning necessarily entails the loss of information. Instead, we proceed via a bootstrap resampling procedure. Consider the case of the mass distribution; distributions of age, extinction, or any combination of the three can be treated in an analogous fashion. Our central estimate for the mass distribution of the population as a whole is simply the sum of the posterior PDFs of mass for all the observed clusters. That is, in a galaxy containing $N$ clusters, for which we have determined a posterior probability distribution for mass $p_i(m)$ for the $i$th cluster, our central estimate of the population PDF is simply
\begin{equation}
p_{\mathrm{pop}}(m) = \frac{1}{N} \sum_{i=1}^{N} p_i(m).
\label{eq:poppdf}
\end{equation}
To estimate the uncertainty on this estimate, we perform $N_t$ trials. For each trial we draw $N$ clusters from our observed sample with replacement, meaning that we may draw the same cluster more than once. We then compute the population PDF $p_{\mathrm{pop},t}(m)$ for that trial via \autoref{eq:poppdf}. Thus after $N_t$ trials we have $N_t$ population PDFs, and at each mass $m$ we have $N_t$ estimates for the value of $p_{\mathrm{pop},t}(m)$. We can then compute a 90\% confidence interval on $p_{\mathrm{pop}}(m)$ as simply the range from the 5th to the 95th percentile of these $N_t$ estimates, and similarly for any other confidence interval of interest to us. The result of this procedure is therefore both a central estimate and a confidence interval on $p_{\mathrm{pop}}(m)$ at each mass $m$, obtained without the need for binning, and using the full posterior PDFs from our Bayesian analysis of the individual clusters.

\subsection{Marginal Mass and Age Distributions}

\begin{figure}
\plotone{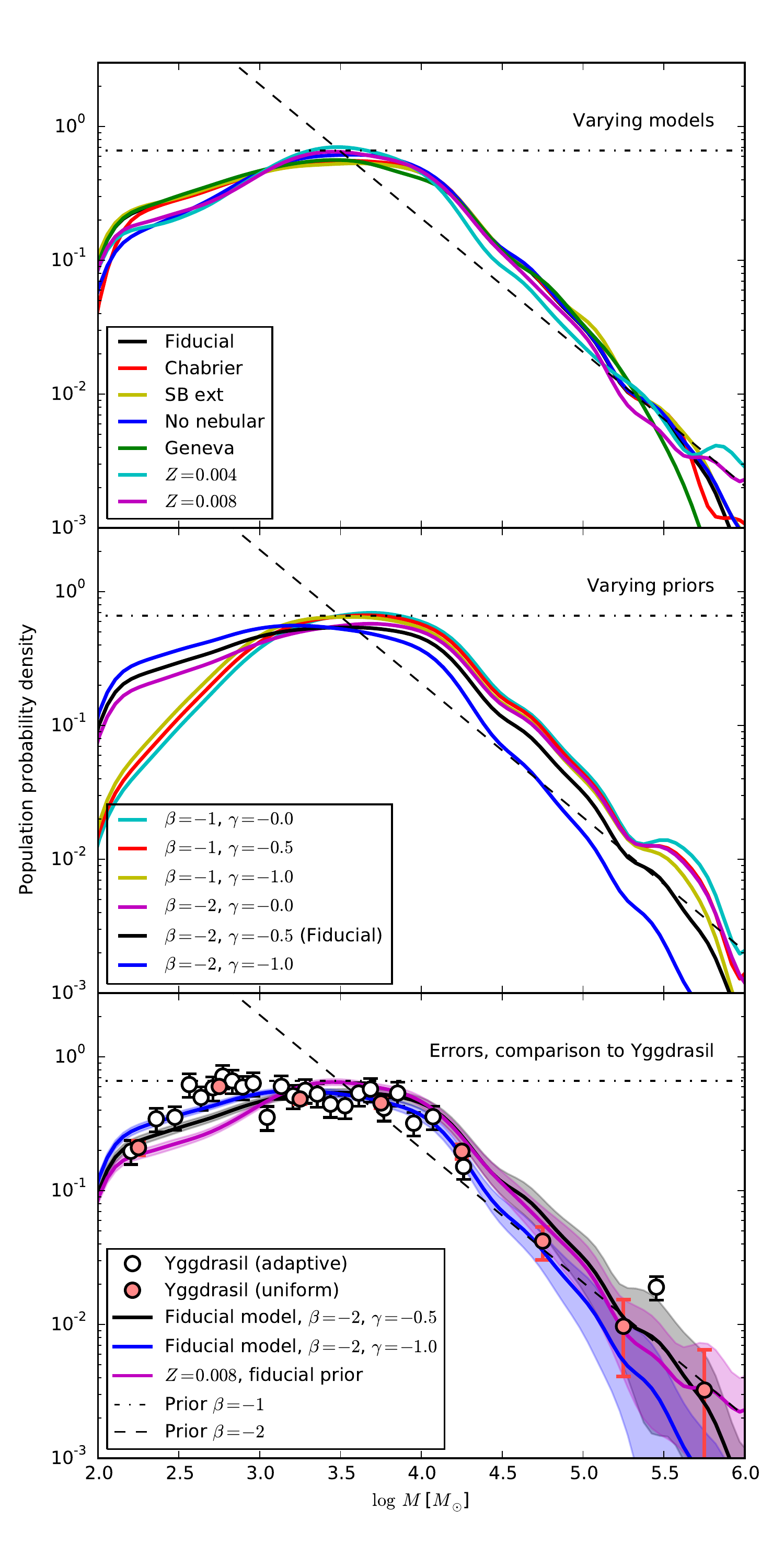}
\caption{
\label{fig:mass_pop}
Inferred posterior PDF for the mass distribution of the entire population of star clusters, marginalized over age and extinction, and summed over the three sample fields. In the top panel, the different colored lines indicate central estimates for this quantity derived from \cs~using the different model libraries: the fiducial one (pad\_020\_kroupa\_MW), one using a Chabrier instead of Kroupa IMF (pad\_020\_chabrier\_MW), one using a starburst instead of Milky Way extinction curve (pad\_020\_kroupa\_SB), one omitting nebular emission (pad\_020\_kroupa\_MW\_noneb), one using Geneva instead of Padova tracks (gen\_020\_kroupa\_MW), and two with metallicities of $Z=0.004$ and $Z=0.008$ instead of $Z=0.020$ (pad\_004\_kroupa\_MW and pad\_008\_kroupa\_MW). In the middle panel, we show central estimates for the mass PDF estimated using our fiducial library (pad\_020\_kroupa\_MW), but a number of different priors $(\beta,\gamma)$, as indicated. In the bottom panel, we repeat some of the models from the top two panels, this time with shaded regions indicating the 5th to 95th percentile confidence interval based on bootstrap resampling with 50,000 trials. Also in the bottom panel, circles with error bars show the values inferred by binning the best fit masses inferred by \yg, using either 25 adaptive bins with equal numbers of clusters per bin (white points), or uniform bins spaced at intervals of $0.5$ dex (salmon points). Error bars on the \yg~points show the Poisson error. The black dashed and dotted lines that appear in all panels show the priors in mass corresponding to $\beta=-2$ and $\beta=-1$.\\
}
\end{figure}

\begin{figure}
\plotone{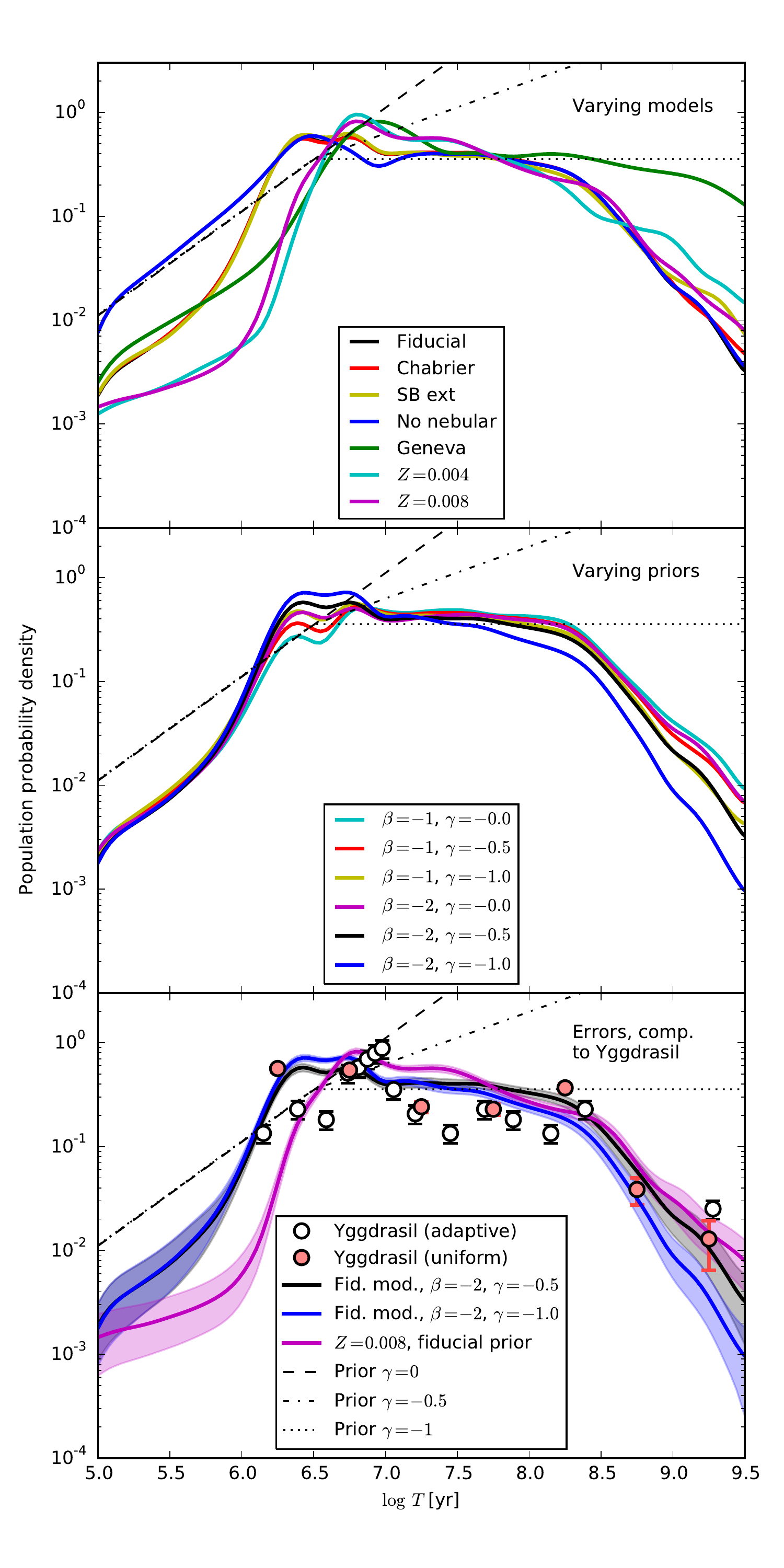}
\caption{
\label{fig:age_pop}
Same as \autoref{fig:mass_pop}, but now showing the age distribution instead of the mass distribution. Black lines show different priors in age corresponding to $\gamma=0$, $-0.5$, and $-1.0$ at ages above $10^{6.5}$ yr.\\
}
\end{figure}

\autoref{fig:mass_pop} shows the inferred marginal mass distributions for the entire population of star clusters, summing over all three of our fields. We emphasize that these plots show $p(\log M)$ rather than $p(M)$, so that a flat line corresponds to equal numbers of cluster per logarithmic mass interval. We see that, for our \cs~based modeling, the mass distribution is consistent with a powerlaw $dN/dM\sim M^{-2}$, or perhaps very slightly steeper, over the mass range from roughly $10^{3.5} - 10^{5.5}$ $M_\odot$. This result is essentially independent of the choice of model library or prior, and with relatively small statistical uncertainty based on bootstrap resampling. At lower masses the mass distribution flattens, almost certainly as a result of incompleteness, and at the lowest masses, $M \lesssim 10^{3}$ $M_\odot$, the inferred posterior distribution depends fairly strongly on the prior. This is a result of posterior PDFs for individual clusters in this mass range being quite broad, so that the total shape depends on the prior. In comparison, the choice of library matters little at low masses. Our central estimate for the shape of the mass function at the very high mass end, approaching $10^6$ $M_\odot$, depends strongly on both the choice of library and the priors, but the statistical uncertainty is so large (due to the small number of clusters) that the bootstrap confidence intervals are mostly overlapping.

\autoref{fig:age_pop} shows the corresponding marginal age distributions. As with \autoref{fig:mass_pop}, we are plotting $p(\log T)$ rather than $p(T)$ (in contrast, for example, to the similar Figure 14 of \citealt{fouesneau14a}), so that a flat line indicates a distribution where the cluster survival probability per decade in age is constant. Absence of cluster disruption would appear as a line of slope $+1$ on this plot.  Based on the \cs~ analysis, these are distributed roughly as $dN/dT \sim T^{-1}$ at ages from roughly $10^7 - 10^{8.5}$ yr, with some dependence of the slope on the choice of prior and on the model library. The largest outlier is the result derived from the Geneva library, but this should probably be regarded as less reliable in the relevant age range due to its omission of TP-AGB stars.

The results with all models and libraries also show an excess of clusters at ages around $10^{6.5}$ yr. This is partly driven by the choice of prior, which is flat in age $T$ (and thus rising in $\log T$) up to this age. We have argued that this is in fact the correct prior to use in this age range, since cluster disruption will be negligible for clusters this young. However, the bump at $10^{6.5}$ yr might also be affected by completeness, which will preferentially suppress clusters at ages older than this due to the rapid decline in luminosity at older ages. At ages much below $10^{6.5}$ yr, the results appear to be quite sensitive to the choice of metallicity and tracks, and the treatment of nebular emission. The feature driving this is almost certainly the variations in nebular emission, which is affected directly by metallicity, and indirectly (though the ionizing luminosity) by the choice of tracks. Eventually all models converge to our prior, $dN/dT \sim \mathrm{constant}$, because, at ages below 1 Myr, even if we include nebular emission then there is no change in cluster colors with age. Since the colors do not distinguish between ages $<1$ Myr, the posterior at such young ages simply reduces to the prior.

We also see a drop off in the cluster population at ages above $\sim 10^{8.5}$ yr. At least some of this drop must occur because these clusters' low luminosities place them below our sensitivity limit, but there may also be a real decline in cluster numbers at older ages as well. If so, this would imply an increase in the rate of cluster disruption at ages around $10^{8.5}$ yr. However, given our incompleteness, it would be premature to draw any conclusions at this point. The shape of this drop off depends somewhat sensitively on the choice of prior, with different but still physically-plausible priors giving shapes in this age range that can be outside each others' 90\% confidence intervals. It does not depend strongly on the choice of library, again with the exception of the Geneva models, which produce quite a different result but are likely less reliable. On the other hand, it is possible that the Padova libraries may also have difficulties in this age range, since it is particularly susceptible to choices in how one treats the hard-to-model TP-AGB phase.

As noted above, to turn the distributions shown in \autoref{fig:mass_pop} and \autoref{fig:age_pop} into mass and age distributions for clusters in general will require completeness correction, a topic that we leave to a future paper. Moreover, the completeness correction will quite different for NGC\,628 than for NGC\,7793 due to their different distances ($\sim 10$ Mpc versus $\sim 3.5$ Mpc). Our goal is therefore not to derive the properties of the cluster populations in these two galaxies, but rather to understand how such determinations are influenced by the method used to convert the photometry into physical properties. We can already see from the results obtained thus far that the age distribution, at least in certain age ranges, will not be independent of the choice of prior, or the choice of library.

Comparing the \cs~results to the ones obtained using \yg, we find that the mass PDFs are roughly consistent within the error bars, though this depends to some extent on how one chooses to bin the \yg~results, and on the prior that one chooses for the \cs~analysis. At small masses where the statistical uncertainties are largest, the \yg~results appear to be closest to what one obtains using \cs~with a prior distribution $\beta=-2$, $\gamma=-1$. For the age distributions there is more discrepancy between the \yg~and \cs~results. The \yg~data show sharp drops in the number of clusters at around $10^{7.3}$ and $10^{8.1}$ yr, with excesses near $10^{6.8}$ and $10^{7.9}$ yr, though the prominence of these features depends on how one chooses to bin. The same structure in the \yg-determined ages is visible as the horizontal banding in \autoref{fig:ygcomp}, and appears to result from the choice to assign a single best fit even in parts of color space where colors provide relatively poor constraints on the true age. The \cs~models correctly capture the uncertainty in such regions by spreading out the posterior probability, and show no corresponding features. Instead, the population PDF is smooth. Nonetheless, the broad distribution of ages in the \yg~models is not dissimilar from that obtained with \cs. Apparently while the stochastic and deterministic methods produce relatively large differences on a cluster-by-cluster basis, they agree much more closely when analyzing the cluster population as a whole.

\subsection{Joint Mass-Age Distribution}

\begin{figure}
\plotone{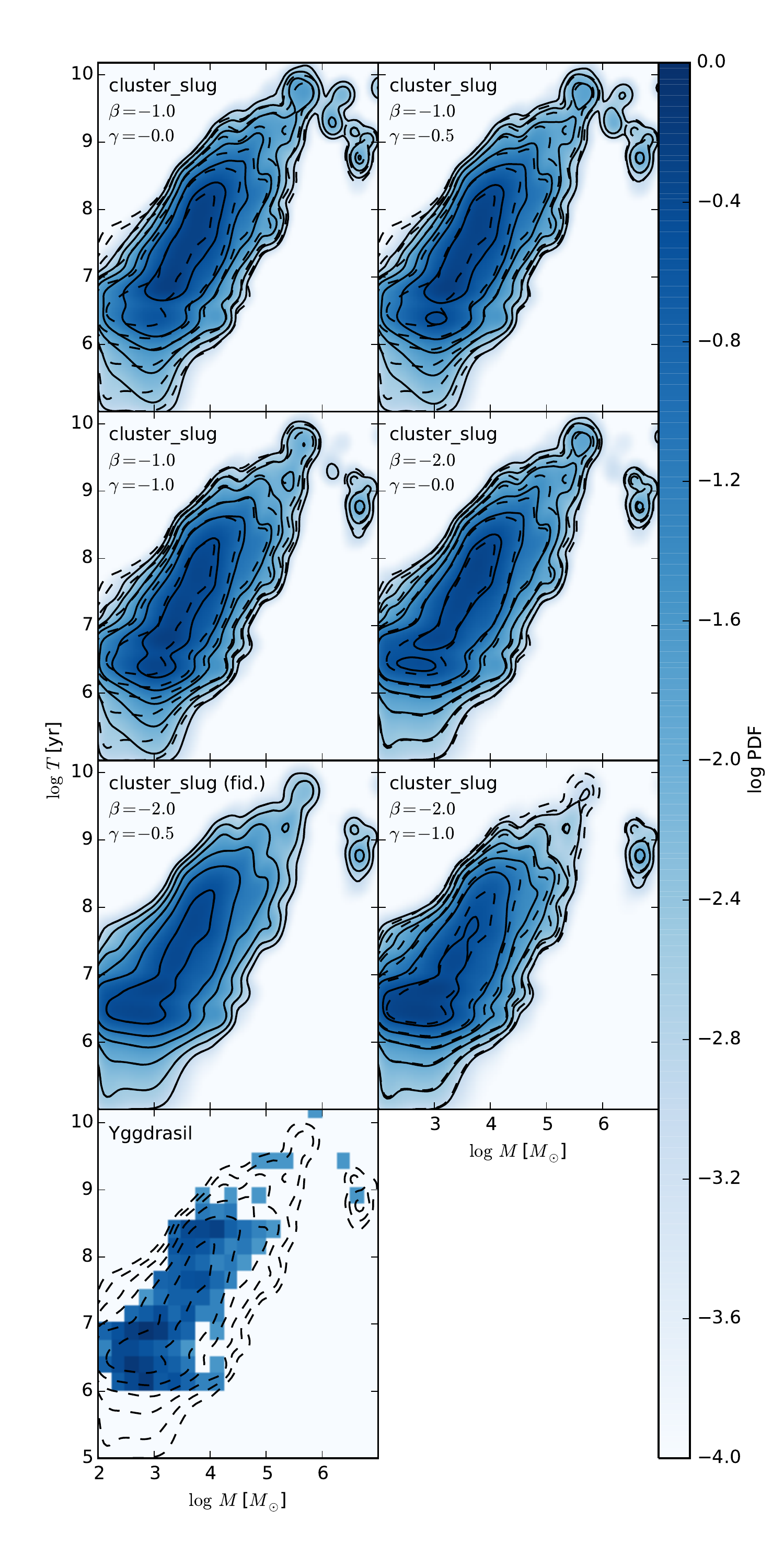}
\caption{
\label{fig:joint_age_mass}
Joint posterior distributions of age and mass for all clusters in all fields, marginalized over extinction. Colors and solid contours show the probability density in the $(\log M, \log T)$ plane; PDFs are properly normalized, so the integral of the PDF over the plane is unity. The top six panels show the results for \cs~using different priors $(\beta,\gamma)$ as indicated, while the bottom panel shows the results for \yg, which have been determined by taking the \yg~best fits and making a 2D histogram using bins $0.25$ dex wide in both the mass and age directions. Dashed contours show the results using \cs~with our fiducial prior ($\beta=-2$, $\gamma=-0.5$), and are the same in every panel in order to facilitate comparison. Both solid and dashed contours start at a log probability density of $-3$, and are spaced at intervals of $0.5$ thereafter.\\
}
\end{figure}

In addition to examining the mass and age distributions separately, we can examine them jointly. \autoref{fig:joint_age_mass} shows the joint distributions of mass and age summed over all three fields, marginalizing over the extinction. All the panels shown use our fiducial model, but we examine a range of priors. We also show the \yg~results, binned in the $(\log M, \log T)$ plane. 

Qualitatively, the \cs~results with all priors and the \yg~results agree that the observed cluster population mostly lies along a band that extends from masses of $\sim 10^3$ $M_\odot$ and ages of $\sim 10^{6.5}$ yr up to masses of $\sim 10^{5.5}$ $M_\odot$ and ages of $\sim 10^{9.5}$ yr. This distribution is doubtless heavily influenced by the completeness of the survey, which would prevent us from detecting clusters that lie in the upper left of the diagram. On the other hand, the absence of clusters in the lower right corner, corresponding to massive, young systems, is a real feature of the distribution, as such clusters should be readily detectable.

Examining the results a bit more closely, we can notice that the choice of priors does influence the final distribution in non-negligible ways. As the priors on mass and age become flatter, $\beta=-1$ versus $\beta=-2$, and $\gamma=0$ versus $\gamma=-1$, clusters tend to shift from the lower left end of the populated band, at low masses and young ages, upward to higher masses and ages. Thus priors that favor low mass and young age, $\beta=-2$, $\gamma=-1$, produce a prominent bump at masses of $10^{2.5-3.5}$ $M_\odot$ and ages near $10^{6.5}$ yr, while flatter priors, $\beta=-1$, $\gamma=0$, remove this feature and distribute the clusters more evenly across a range of masses and ages. Flatter priors also produce a small island of probability at a mass of $\sim 10^{6.5}$ $M_\odot$ and an age of $\sim 10^{9.5}$ yr. This island is produced by a handful clusters that are not well fit by either the \cs~or~\yg~models, making the \cs~results for them quite sensitive to the prior.\footnote{For example, the cluster for which \yg~assigns the highest mass, which appears in the right-most pixel in the \yg~panel of \autoref{fig:joint_age_mass}, has a $Q$ parameter of $1.9\times 10^{-4}$, indicating a very poor fit. The normalized photometric nearest-neighbor distance $D_1^{\mathrm{norm}} = 3.0$ for our fiducial library, indicating that no \cs~models land closer than $3\sigma$ to the cluster's photometric properties; alternate libraries do no better. Given the strong disagreement with all sets of models, it seems likely that this object is not well described as a simple stellar population, or that all the model tracks we have available are deficient.}

The other very noticeable difference is between the \yg~and \cs~models. Here the key feature is that the \yg-derived masses and ages are confined to a much smaller portion of the age-mass plane than the \cs~ones. This reflects the much smaller uncertainties that \yg~assigns based on its $\chi^2$ fitting approach, as opposed to the Bayesian method of \cs. This approach tends to concentrate the probability toward the peaks, suppressing the lower probability wings that are retained by using the full posterior.

\section{Summary and Conclusions}
\label{sec:conclusions}

We describe and investigate the performance of a Bayesian method to derive the masses, ages, and extinctions of star clusters observed using broadband photometry. Our sample data set for this analysis consists of 621 visually-confirmed star clusters drawn from NGC\,628e and NGC\,7793. These galaxies are part of the Legacy Extragalactic UV Survey (LEGUS) sample, which provides \textit{HST} photometry in the NUV, U, B, V, and I bands. Our method, implemented based on the \slug~software suite\footnote{\url{https://www.slugsps.com}} and its cluster analysis tool \cs, uses kernel density estimation coupled to implied conditional regression to return the full marginal posterior probability distributions of mass, age, and extinction. This technique proves to be particularly useful in two regimes. One is for low-mass clusters, below $\sim 10^{3.5}$ $M_\odot$, where the initial mass function is not fully sampled and the relationship between physical and photometric properties is therefore non-deterministic, leading a broad posterior PDF that is not well characterized by a single best fit. The second case is for clusters that lie at locations in color space where there are significant degeneracies between mass, age, and extinction, so that there are multiple possible fits of comparable likelihood at distinct sets of physical parameters. In both of these regimes, a full posterior PDF provides a more reliable result, and a more realistic estimate of the errors, than methods that return a single best fit with a local error distribution around it. Indeed, by comparing our results to those produced by the non-stochastic cluster fitting code \yg, we find that the stochastic method removes some of the artifacts found in the deterministic one, and that it returns substantially larger error distributions.

At the level of individual clusters, our analysis technique proves to be insensitive, within the plausible range of variations, to our choice of metallicity, extinction curve, stellar initial mass function, and evolutionary tracks. On the other hand, because in many cases the observed photometry is consistent with more than one combination of mass, age, and extinction, our choice of prior probabilities does affect the final result. This sensitivity occurs because, when more than one mass-age-extinction combination is consistent with the observed photometry, the relative weight we assign to different ``islands" of probability depends on our priors. This means that, for an individual cluster, before drawing any firm conclusions about its age, or to a less extent its mass, one should be careful to ensure that the results are robust against variations in priors, whether explicit or implicit. It is possible that the addition of extra photometric data, particularly H$\alpha$, might break some of these degeneracies, but that remains to be seen.

At the level of a cluster population, we find that the mass distribution in the range $\sim 10^3-10^6$ $M_\odot$ is very robust against changes in either prior or assumed stellar models, and to whether we derive the masses using stochastic \cs~models or deterministic \yg~ones. At lower masses, priors begin to matter because the large amount of stochastic variation induced by incomplete IMF sampling means that photometry no longer provides strong constraints on mass, leaving the posterior close to the prior. At the highest masses stochasticity is unimportant, but depending on the assumed metallicity and choice of tracks, the inferred mass can vary by up to $\sim 0.5$ dex. When the distribution is dominated by a small number of clusters, this translates directly into variations in the overall distribution. Age distributions are somewhat less robust that mass ones. They are at least mildly dependent on the choice of prior at all ages, and at ages $\lesssim 3$ Myr when nebular emission contributes significantly to the light, they are very sensitive to metallicity, choice of tracks, and the assumed efficiency with which ionizing photons are converted to nebular light within the observational aperture. They are also quite dependent on the choice of prior at young ages.

The method we develop in this paper is executed in an automated pipeline, which is efficient enough that we can derive marginal posterior PDFs for the high confidence cluster catalog of $621$ clusters in well under an hour on a workstation-level machine. This pipeline is available at \url{https://www.slugsps.com}, and will form the basis of a full stochastic cluster catalog derived from all the star clusters in the LEGUS sample. These will be expanded to include H$\alpha$ data from the H$\alpha$-LEGUS program (PI: R.~Chandar) as they become available. The resulting data set will provide an unprecedented sample of cluster properties, with realistic uncertainty distributions, from across the star-forming part of the Hubble sequence.

\acknowledgements Based on observations made with the NASA/ESA Hubble Space Telescope, obtained at the Space Telescope Science Institute, which is operated by the Association of Universities for Research in Astronomy, Inc., under NASA contract NAS 5-26555. These observations are associated with program \#13364. Support for program \#13364 was provided by NASA through a grant from the Space Telescope Science Institute. MRK acknowledges funding from HST theoretical research program \#13256, which provided support for the development of the \slug~code. MF acknowledges support by the Science and Technology Facilities Council, grant number ST/L00075X/1. DAG kindly acknowledges financial support by the German Research Foundation (DFG) through grant GO 1659/3-2. EZ acknowledges research funding from the Swedish Research Council (project 2011-5349).

\bibliographystyle{apj}
\bibliography{refs}

\end{document}